\documentclass[
    aps,
    prd,
    reprint,
    superscriptaddress,
    nofootinbib
]{revtex4-2}

\usepackage{amssymb}
\usepackage{amsfonts}
\usepackage{prodint}
\usepackage{graphicx}
\usepackage{bbold}
\usepackage{upgreek}
\usepackage{dsfont}
\usepackage{mathrsfs}
\usepackage{xcolor}
\usepackage{comment}
\usepackage{slashed}
\usepackage{etoolbox}
\usepackage{braket}
\usepackage{mathtools}
\usepackage{tikz}
\usetikzlibrary{arrows.meta,decorations.pathreplacing}
\usepackage{times}

\usepackage[
    colorlinks=true,
    citecolor=teal,
    linkcolor=blue,
    urlcolor=teal
]{hyperref}

\usepackage{orcidlink}

\begin{document}

\title{Quantization of the electroweak Volkov--Proca theory}
\author{Tudor P\u{a}tuleanu~\orcidlink{0000-0001-8594-4606}}
\email[Corresponding author: ]{tudor.patuleanu@e-uvt.ro}
\affiliation{Department of Physics, West University of Timi\c{s}oara, Bd.~Vasile P\^arvan 4, Timi\c{s}oara 300223, Rom\^ania}
\affiliation{Institut Denis Poisson, CNRS UMR 7013, Universit\'e de Tours, Universit\'e d'Orl\'eans, 
Parc de Grandmont, Tours, 37200, France}

\begin{abstract}

We quantize the charged sector of the electroweak theory in an external electromagnetic plane-wave background after spontaneous symmetry breaking, which we refer to as the electroweak Volkov--Proca theory, and analyze its global $U(1)$ symmetry and continuous residual translation subgroup together with the associated conserved charges and momentum components that label the states. We further derive the completeness and orthonormality relations for the Volkov--Proca states in Ritus form, and construct the off-shell and on-shell propagators, including their representation in terms of the scalar Volkov propagator. Finally, we derive the K\"all\'en--Lehmann spectral representation and the LSZ reduction formula for the Volkov--Proca states.

\end{abstract}

\maketitle

\section{Introduction}
\label{sec: introduction}

There has been sustained interest in quantum electrodynamics processes in strong external electromagnetic fields, sparked by the seminal works of A.~I.~Nikishov \cite{Nikishov:1964zza} and V.~I.~Ritus \cite{Ritus:1985vta, Ritus1972} and V.~N.~Baier and V.~M.~Katkov \cite{Baier:1975ys, Baier:1998vh}, with continued theoretical developments in more recent studies \cite{Meuren:2013oya, DiPiazza:2018ofz} and the proposal of dedicated strong-field experiments \cite{Fedotov:2022ely}. Several electromagnetic background field configurations, characterized in part by the two Lorentz invariants of the electromagnetic field, $\mathfrak F \equiv \frac{1}{4}F_{\mu\nu}F^{\mu\nu} = \frac{1}{2}\left(\boldsymbol B^2-\boldsymbol E^2\right)$ and $\mathfrak G \equiv \frac{1}{4}F_{\mu\nu}\widetilde F^{\mu\nu} = -\boldsymbol E\cdot\boldsymbol B$, have been considered in the analysis of radiative processes. In particular, constant magnetic fields, for which $\mathfrak F=B^2/2>0$ and $\mathfrak G=0$, \cite{Adhikari:2024bfa, PhysRevD.96.076014, Piccinellis:2017aa, PhysRevD.97.103008, PhysRevD.99.056011, PhysRevD.107.116024, Piccinelli:2023ab, Ayala:2026dgh, Ayala:2024ucr, Ayala:2026eja, Ayala:2026nte}, constant crossed fields, satisfying $\mathfrak F=\mathfrak G=0$, \cite{Ritus1972, King:2013, King:2018, Mironov:2020}, and electromagnetic plane-wave backgrounds, which likewise satisfy $\mathfrak F=\mathfrak G=0$, have been extensively investigated. 


In contrast, electroweak processes in strong field backgrounds have received considerably less attention, particularly even less for plane-wave configurations. However, the tools and ingredients needed to carry out calculations involving the charged electroweak sector have been laid out in other configurations of electromagnetic backgrounds. For example, the charged scalar, fermion and massive vector boson propagators have been calculated, given a constant and homogeneous magnetic background field, the latter in a variety of gauge and metric choices, and different representations \cite{EigenfunctionMethod_ritus, ElectroweakProcesses_kuznetsov,ChargedMassive_iablokov, Elizalde:2000vz, Erdas:2000iq, PhysRevD.108.016012}. More recently, Ref.~\cite{Monreal5jyf-llwf}
constructed the propagator of a massive charged vector boson in the
unitary gauge using both the Ritus eigenfunction method and the Schwinger
proper-time formalism, and derived a Lehmann-Symanzik-Zimmermann (LSZ) reduction formalism based on dressed asymptotic states for the consistent treatment of external charged particles in the magnetic background. 

For electromagnetic plane-wave backgrounds, propagators for charged scalar, fermionic, and massive vector fields have also been obtained, for example in Ref.~\cite{Kurilin:1999qc}, although not within the Ritus eigenfunction formalism. An important ingredient of the Ritus approach is the establishment of completeness and orthonormality relations for the Ritus matrices and the corresponding Volkov states. For charged fermions in a plane-wave background, such relations were formulated in Ref.~\cite{DiPiazza:2018ofz}, building on earlier work from Ref.~\cite{Boca:2011gvq}. 

In the present work, we extend these results to the charged massive vector bosons arising from electroweak theory after spontaneous symmetry breaking, working with the physical fields corresponding to the unitary gauge. Specifically, in Sec.~\ref{sec: Volkov-Proca_theory}, we formulate the
electroweak Volkov--Proca theory in an external electromagnetic plane-wave
background and discuss its global $U(1)$ and residual translation symmetries together with the associated conserved quantities. In Sec.~\ref{sec: Volkov-Proca_states}, we recast the Volkov--Proca solution of Ref.~\cite{Brown:1983bc} in Ritus form, establish its spectral properties, and show that the spectral momentum remains supported on the standard mass-shell hyperboloid, $p^2=M_W^2$. This differs from the case
of a constant magnetic field, where the Landau-level structure modifies
the dispersion relation to $E^{2}=M_W^{2}+(p^{3})^{2}+2|eB|(\ell-1/2)$
\cite{Monreal5jyf-llwf}. In Sec.~\ref{sec: completeness_orthonormality}, we derive the completeness relations for the dressed polarizations and establish the completeness and orthonormality relations for the Volkov--Proca Ritus matrices, including the corresponding off-shell Proca-numerator identity, while
the on-shell Volkov--Proca modes satisfy the associated indefinite
sesquilinear orthonormality relations. In Sec.~\ref{sec: Volkov-Proca_propagator}, we solve the Green equation in the Ritus representation and construct the off-shell and on-shell forms of the advanced, retarded, and Feynman Volkov--Proca propagators. In Sec.~\ref{sec: quantization}, we canonically quantize the theory, introduce the creation and annihilation operators, express the propagators in terms of Wightman functions and the projectors onto the positive- and negative-frequency sectors, and quantize the conserved charges, showing that the charges associated with the global $U(1)$ symmetry and continuous residual translation symmetries admit diagonal representations. In Sec.~\ref{sec: LSZ_reduction}, we derive the LSZ reduction formula for charged vector bosons in an electromagnetic plane-wave background, providing the formalism required for the consistent treatment of background-dressed external $W$-boson states in scattering amplitudes. 
Finally, taking into account the reduced translation symmetry of the plane-wave background, in Sec.~\ref{sec: Kallen-Lehmann_spectral} we construct the corresponding K\"all\'en--Lehmann spectral representation, while in
Sec.~\ref{sec: phase_evolution} we derive the phase-evolution operator
governing the remaining dependence of the Volkov--Proca modes on the
plane-wave phase.

\section{Volkov--Proca theory}
\label{sec: Volkov-Proca_theory}

On Minkowski spacetime $\eta^{\mu\nu} = \operatorname{diag} \left(1, -1, -1, -1\right)$, consider an external plane-wave electromagnetic background field (Volkov background \cite{Volkov1937}), provided by
\begin{equation}
    \mathcal{A}^\mu_\mathrm{bg} = e A^\mu (\phi),
\end{equation}
where $e$ denotes the electric charge and $A^\mu$ is the background vector potential, which depends only on the plane-wave phase $\phi \equiv x^+ = n \cdot x$, with $n^\mu$ the null wave vector,
$n^2=0$, as detailed in Eqs.~\eqref{eq: light_cone_basis} and
\eqref{eq: light_cone_coordinates} of the Appendix.

The Lorenz gauge condition is adopted \cite{Berestetskii:1982qgu},
\begin{equation}
    \partial_\mu \mathcal{A}^\mu_\mathrm{bg} 
    = n \cdot \mathcal{A}^\prime_\mathrm{bg} 
    = (n \cdot \mathcal{A}_\mathrm{bg} )^\prime
    = 0,
\end{equation}
which, since $\partial_\mu = n_\mu \partial_\phi$, implies that
$\mathcal{A}^+ \equiv n \cdot \mathcal{A}$ is constant, which can be set to zero under the residual gauge freedom $\Lambda(\tilde{\phi})$, with $\tilde{\phi} \equiv x^- = \tilde{n} \cdot x$, where $\tilde{n}$ is the second null ($\tilde{n}^2 = 0$) vector of the light-front basis with $n \cdot \tilde{n} = 1$. Another residual gauge transformation, as explained in Appendix \ref{app: gauge_choice}, allows to set $\mathcal{A}^- \equiv \tilde{n} \cdot \mathcal{A} = 0$, and to consider only perpendicular polarizations. In turn, as shown before \eqref{eq: gauge_choice}, this implies that $\mathcal{A}^0 = 0$, which will be used throughout.

\subsection{Lagrangian and equations of motion}

The charged sector of electroweak theory, after spontaneous symmetry breaking, in an electromagnetic plane-wave background, is described, in vacuum (turning off all fields except those of interest) by the (electroweak) Volkov-Proca ($\mathrm{VP}$) Lagrangian,
\begin{equation}
    \label{eq: Volkov-Proca_Lagrangian}
    \begin{aligned}
    \mathcal{L}_\mathrm{VP} = 
    &- \frac{1}{2} \mathcal{W}^{\mathrm{bg}\, +}_{\mu\nu} \mathcal{W}^{-\, \mu\nu}_\mathrm{bg} + M_W^2 W^+_\mu W^{-\,\mu}\\
    &+ i \mathcal{F}^{\mu\nu}_\mathrm{bg} W^+_\mu W^-_\nu,
    \end{aligned}
\end{equation}
which includes the additional term $i \mathcal{F}^{\mu\nu}_\mathrm{bg} W^+_\mu W^-_\nu$ to the minimally coupled Proca Lagrangian \cite{Proca:1936fbw}. Here, the interaction with the electromagnetic background is through the electroweak field strength tensor
\begin{equation}
    \mathcal{W}^{\mathrm{bg}\, \pm}_{\mu\nu} = D^{\mathrm{bg}}_\mu W^\pm_\nu - D^{\mathrm{bg}}_\nu W^\pm_\mu,
\end{equation}
with the background covariant derivative
\begin{equation}
    (D_Q^{\mathrm{bg}})_\mu \equiv (D^{\mathrm{bg}}_Q)_\mu (\phi) = \partial_\mu + i Q \mathcal{A}_{\mu}^{\mathrm{bg}} (\phi),
\end{equation}
where $Q$ is the charge-sector operator, acting as
\begin{equation*}
    Q W_q^\mu = q W_q^\mu,
    \qquad
    q = \pm 1,
\end{equation*}
and also through the additional term that involves 
\begin{equation}
\label{eq: electromagnetic_background_field_strength_tensor}
\begin{aligned}
    \mathcal{F}_{\mathrm{bg}}^{\mu\nu} 
    &= \partial_\mu \mathcal{A}^\nu_{\mathrm{bg}} - \partial_\nu \mathcal{A}^\mu_{\mathrm{bg}} = e F_{\mathrm{bg}}^{\mu\nu} (\phi)\\
    &= n_\mu \mathcal{A}'^\nu_{\mathrm{bg}} - n_\nu \mathcal{A}'^\mu_{\mathrm{bg}},
\end{aligned}
\end{equation}
the charge-multiplied electromagnetic field strength tensor.

Dropping the ``$\mathrm{bg}$'' subscripts and superscripts in the notation,
we introduce the dressed momentum operator
\begin{equation}
    \label{def: momentum_operator}
    \Pi_Q^\mu
    \equiv
    iD_Q^\mu
    =
    i\partial^\mu
    -
    Q\mathcal A^\mu(\phi),
    \qquad
    P^\mu
    \equiv
    i\partial^\mu.
\end{equation}

The equation of motion (EOM) operator is defined as
\begin{equation}
    \label{def: VP_EOM_operator}
    \left(\mathcal O_Q\right)^\mu{}_\nu
    \equiv
    \left(\Pi_Q^2-M_W^2\right)\delta^\mu{}_\nu
    -
    \Pi_Q^\mu\Pi_{Q\,\nu}
    -
    2iQ\mathcal F^\mu{}_\nu,
\end{equation}
such that the Euler--Lagrange equation reads compactly
\begin{equation}
    \label{eq: VP_EOM_full}
    \left(\mathcal O_q\right)^\mu{}_\nu
    W^\nu_q (x)
    =
    0,
\end{equation}
where the charge sector operator $Q$ has been replaced by $q$. 

Acting with another dressed momentum operator leads to 
\begin{equation} 
\label{eq: transversality_constraint} 
\Pi^\mu_q W^q_\mu (x) = 0, 
\end{equation} 
a transversality constraint that simplifies the EOM to \cite{Monreal5jyf-llwf} 
\begin{equation} 
\label{eq: EOM} 
\left[(\Pi^2_q - M_W^2) \delta^{\mu}_{\, \nu} - 2 i q \mathcal{F}^{\mu}_{\, \nu}\right] W^\nu_q (x) = 0.
\end{equation}

\subsection{Symmetries of the theory}

\subsubsection{Global \texorpdfstring{$U(1)$}{U(1)} symmetry}

The global $U(1)$ symmetry of the Lagrangian in
Eq.~\eqref{eq: Volkov-Proca_Lagrangian} gives rise to the conserved on-shell Noether current, 
\begin{equation}
    j^\mu
    =
    - i\left(
        \mathcal W^{+\,\mu\nu}W^-_\nu
        -
        \mathcal W^{-\,\mu\nu}W^+_\nu
    \right),
    \qquad
    \partial_\mu j^\mu=0.
\end{equation}
leading to a (time-)conserved charge
\begin{equation}
    \label{eq: conserved_charge_classical}
    \mathcal Q_{\mathrm{el}} = e \int d^3 x\, j^0 (x), \qquad \frac{d}{d t} \mathcal Q_{\mathrm{el}} = 0,
\end{equation}
in analogy to the free complex Proca field from \cite{Radu:2017ufc, Cotaescu:2026}. The sesquilinear current form between two solutions $U$ and $V$ belonging to the same charge sector $q$ is then
\begin{equation}
j^\mu(U,V)
=
i\left(
\bar{\mathcal W}^{\mu\nu}[U]\,V_\nu
-
\mathcal W^{\mu\nu}[V]\,\bar U_\nu
\right),
\end{equation}
where the field strength tensor associated to a solution is
\begin{equation}
\begin{aligned}
    \mathcal W_q^{\mu\nu}[V]
    &=
    -i\left(
        \Pi_q^\mu V^\nu
        -
        \Pi_q^\nu V^\mu
    \right),\\
    \bar{\mathcal W}_{-q}^{\mu\nu}[U]
    &=
    -i\left(
        \Pi_{-q}^\mu \bar U^\nu
        -
        \Pi_{-q}^\nu \bar U^\mu
    \right),
\end{aligned}
\end{equation}
with the corresponding complex-conjugate field strength.

Since the products from $j^\mu(U,V)$ are electrically neutral,
\begin{equation}
    \partial_\mu\left(X_{-q}Y_q\right)
    =
    -i \left[
        \left(\Pi_{-q,\mu}X_{-q}\right)Y_q
        +
        X_{-q}\left(\Pi_{q,\mu}Y_q\right)
    \right],
    \label{eq: neutral_product_covariant_derivative}
\end{equation}
this leads to the on-shell vanishing divergence,
\begin{equation}
    \partial_\mu j^\mu(U,V)
    =
    \left(
        \Pi_{-q,\mu}
        \bar{\mathcal W}_{-q}^{\mu\nu}[U]
    \right)V_\nu
    -
    \bar U_\nu
    \Pi_{q,\mu}
    \mathcal W_q^{\mu\nu}[V].
    \label{eq: current_divergence_momentum}
\end{equation}
Equivalently, in terms of the EOM operator defined in
Eq.~\eqref{def: VP_EOM_operator},
\begin{equation}
    \partial_\mu j^\mu(U,V)
    =
    i\left[
        \bar U_\mu
        (\mathcal O_q)^\mu{}_\nu
        V^\nu
        -
        \bar U_\mu
        (\overleftarrow{\mathcal O}_{-q})^\mu{}_\nu
        V^\nu
    \right].
    \label{eq: current_divergence_EOM}
\end{equation}

The conserved current induces the sesquilinear form
\begin{equation} 
(U,V) = \int_{\Sigma} d\Sigma_\alpha\, j^\alpha(U,V),
\end{equation}
integrated over a spatial Cauchy hypersurface. Choosing
\begin{equation}
    \Sigma_t
    =
    \{x^\mu\in\mathbb{R}^{1,3}\mid x^0=t\},
\end{equation}
with normal vector $u^\alpha=(1, \boldsymbol 0)$, this becomes
\begin{equation}
    (U,V)
    =
    \int d^3x\,j^0(U,V).
\end{equation}

To establish the hypersurface independence of this form, consider the spacetime slab
\begin{equation}
    \Omega_{12}
    =
    \left\{
        x^\mu=(t,\boldsymbol x)
        \,\middle|\,
        t_1\leq t\leq t_2
    \right\}.
\end{equation}
Integrating the divergence of the current over $\Omega_{12}$ gives
\begin{equation}
\begin{aligned}
    \int_{\Omega_{12}} d^4x\,
    \partial_\mu j^\mu(U,V)
    &=
    \int_{t_1}^{t_2}dt
    \int d^3x\,
    \left(
        \partial_0 j^0
        +
        \partial_i j^i
    \right)
    \\
    &=
    (U,V)_{t_2}
    -
    (U,V)_{t_1}\\
    &+
    \int_{t_1}^{t_2}dt
    \oint_{\partial\Sigma_t}
    dS_i\,j^i(U,V).
\end{aligned}
\end{equation}
For localized fields, the boundary term vanishes,
\begin{equation}
    \int_{\Omega_{12}} d^4x\,
    \partial_\mu j^\mu(U,V)
    =
    (U,V)_{t_2}
    -
    (U,V)_{t_1}.
\end{equation}
For two on-shell solutions, current conservation implies
\begin{equation}
    \label{eq: inner_product_time_independence}
    \frac{d}{dt}(U,V)=0.
\end{equation}

If the external background vanishes asymptotically, the conserved sesquilinear product evaluated in the free asymptotic region has the same value at all intermediate times.

Using a gauge for which $\mathcal A^0=0$, together with the
transversality constraint and integration by parts, and taking the
Volkov--Proca states to be suitably localized wave packets such that the
surface terms vanish at spatial infinity, the sesquilinear form reduces
to
\begin{equation} 
\label{eq: inner_product}
(U,V) = - i\int d^3x\, \bar U_\mu \overleftrightarrow{\partial}^{\,0} V^\mu, 
\end{equation}
where the bidirectional derivative is defined as
\begin{equation}
    \bar U_\mu
    \overleftrightarrow{\partial}^{\,0}
    V^\mu
    \equiv
    \bar U_\mu\,\partial^0 V^\mu
    -
    \left(\partial^0\bar U_\mu\right)V^\mu.
\end{equation}
This is the same expression as for the free theory from \cite{Cotaescu:2026}.

\subsubsection{Spacetime symmetry}

Although the external plane-wave background breaks the full Poincaré symmetry, in the following we restrict attention to its continuous residual translation subgroup. Its mutually commuting generators provide the conserved momentum labels used to classify the Volkov--Proca states and determine the corresponding momentum-conservation laws in scattering processes. These translations, 
\[
x^\mu \to x^\mu+\xi^\mu,
\]
leave the phase $\phi = n \cdot x$ unchanged provided
\begin{equation}
\phi
\to
\phi+n\cdot\xi
\stackrel{!}{=}
\phi
\quad\Longrightarrow\quad
\xi\cdot n=0.
\end{equation}
Under this condition, the infinitesimal variation
\begin{equation}
    \delta_\xi \mathcal{A}^\mu = \xi \cdot \partial \mathcal A^\mu(\phi)
    =
    (\xi\cdot n)\mathcal A^{\prime\mu}(\phi)
    =
    0.
\end{equation}

For a periodic background of period $T$,
\begin{equation}
    \mathcal A^\mu(\phi+T)
    =
    \mathcal A^\mu(\phi),
\end{equation}
the translation symmetry is enlarged by discrete transformations that shift the phase by an integer number of periods,
\begin{equation}
    n\cdot\xi=mT,
    \qquad
    m\in\mathbb Z.
\end{equation}

The vector potential is therefore invariant under both the continuous translations with $m=0$, which form the component connected to the identity, and the additional discrete translations with $m\neq0$, of the residual translation group
\begin{equation}
    \label{eq: residual_translation_group}
    T_{\mathcal A}
    =
    \left\{
        \xi^\mu\in\mathbb R^{1,3}
        \,\middle|\,
        n\cdot\xi\in T\mathbb Z
    \right\}
    \simeq
    \mathbb R^3\times\mathbb Z,
\end{equation}

The continuous part of the residual translation symmetry,
\begin{equation}
    \label{eq: residual_connected_translation_group}
    T_{\mathcal A}^{(0)}
    =
    \left\{
        \xi^\mu
        \,\middle|\,
        n\cdot\xi=0
    \right\}
    \simeq
    \operatorname{span}\{n,a_1,a_2\}
    \simeq
    \mathbb R^3,
\end{equation}
gives rise, through Noether's theorem, to conserved currents
\begin{equation}
    J_\xi^\rho
    =
    T_{\mathrm{can}}^{\rho\mu}\xi_\mu,
    \qquad
    \partial_\rho J_\xi^\rho=0, 
    \qquad \xi \cdot n = 0,
\end{equation}
with the associated charge, quantized in Subsec~\ref{subsec: conserved_charges},
\begin{equation}
    \label{eq: conserved_charge_phase_translation}
    \mathcal Q_\xi
    =
    \int d^3x\,J_\xi^0
    =
    \xi \cdot P_{\mathrm{can}},
    \qquad
    \frac{d\mathcal Q_\xi}{dt}=0,
\end{equation}
where the canonical four-momentum is defined by
\begin{equation}
    P_{\mathrm{can}}^\mu
    =
    \int d^3x\,T_{\mathrm{can}}^{0\mu},
\end{equation}
in terms of the canonical energy-momentum tensor
\begin{equation}
    \label{eq: canonical_energy_momentum_tensor}
    T_{\mathrm{can}}^{\rho\mu}
    =
    \sum_{q = \pm}
    \mathsf\Pi_{- q}^{\rho\alpha}
    \partial^\mu W^{q}_{\alpha}
    -
    \eta^{\rho\mu}\mathcal L_{\mathrm{VP}},
\end{equation}
where the canonical momentum tensor is\footnote{The subscript $q$ denotes the charge sector of the corresponding field.} $\mathsf\Pi_{q}^{\rho\alpha} = - \mathcal W_q^{\rho\alpha}$.

Since the residual continuous translation group is Abelian,
\begin{equation}
    [
        \mathcal Q_\xi,
        \mathcal Q_\eta
    ]
    =
    0,
    \qquad
    \xi\cdot n
    =
    \eta\cdot n
    =
    0,
\end{equation}
the fact that the generators are commuting\footnote{After quantization, or at the classical level in terms of Poisson brackets.} means that they can be simultaneously diagonalized, allowing the states to be labelled by their corresponding eigenvalues, as in Sec.~\ref{sec: Kallen-Lehmann_spectral}.

\section{Volkov--Proca states}
\label{sec: Volkov-Proca_states}

Expanding the field in terms of momentum modes, leads to the equation of motion for each mode,
\begin{equation}
    \left[(\Pi_q^2 - M_W^2) \delta^{\mu}_{\, \nu} - 2 i q \mathcal{F}^{\mu}_{\, \nu}\right] w^\nu_q (p, x) = 0,
\end{equation}
with the Ritus ansatz \cite{Ritus:1985vta, Monreal5jyf-llwf},
\begin{equation}
    \label{eq: EOM_mode_soln_off_shell}
    w_q^\mu(p,x)
    =
    E_q (p, x)^\mu_{~\,\nu}\,\chi^\nu(p),
\end{equation}
where $\chi^\nu(p)$ denotes the polarization vector associated with the spectral momentum $p$, reducing on shell to the usual free Proca polarization vector.

\subsection{Classical covariant Volkov--Proca states}

The covariant solution found in \cite{Brown:1983bc} has a Ritus matrix
\begin{equation} \label{eq: RitusMatrix}
\begin{aligned}
    E_q (p, x)^\mu_{~\,\nu} 
    &\eqqcolon e^{- i p \cdot x} E_q (p, \phi)^\mu_{~\,\nu},
\end{aligned}
\end{equation}
in terms of the phase-dependent Ritus matrix
\begin{equation}
    E_q(p,\phi)^\mu{}_\nu
    \coloneqq
    U_q(p,\phi)\,
    \Lambda_q(p,\phi)^\mu{}_\nu,
    \label{eq: phase_dependent_Ritus_matrix}
\end{equation}
where the Volkov phase factor is
\begin{equation}
    U_{q} (p, \phi) = \exp \left\{- i S_q (p, \phi)\right\},
\end{equation}
expressed in terms of the Ritus phase
\begin{equation}
\begin{aligned}
    S_{q} (p, \phi) 
    &\coloneqq \frac{q}{(n \cdot p)} p \cdot \int^\phi_{\phi_0} d\varphi\, \mathcal{A} (\varphi)\\ 
    &- \frac{1}{2 (n \cdot p)} \int^\phi_{\phi_0} d\varphi\, \mathcal{A}^2 (\varphi),
\end{aligned}
\end{equation}
where $\phi_0$ is a fixed reference phase, which may be chosen $\phi_0=-\infty$ whenever the background admits an asymptotic past in which the corresponding integrals are well defined. The Lorentz transformation from the little group\footnote{It leaves the wave vector $n^\mu$ invariant, that is, $\Lambda \cdot n = n$.} of $n^\mu$,
\begin{equation}
\Lambda_q(p,\phi)^\mu{}_\nu
\coloneqq
\left[
\exp\!\left(
    \frac{q}{n\cdot p}\,
    N(\phi)
\right)
\right]^{\mu}{}_{\raisebox{-1.4ex}{$\scriptstyle \nu$}},
\end{equation}
where the phase-dependent null-rotation generator is
\begin{equation}
\label{eq: phase_dependent_null_rotation_generator}
N(\phi)^\mu{}_\nu \coloneqq n^\mu \mathcal A_\nu (\phi) - \mathcal A^\mu (\phi) n_\nu.
\end{equation}

The transversality constraint in Eq.~\eqref{eq: transversality_constraint}, together with the dressing relations in Eq.~\eqref{eq: momentum_dressing_undressing} of App~\ref{app: momentum_dressing}, is equivalent to
\begin{equation}
    p\cdot\chi(p)=0,
\end{equation}
showing that $\chi^\mu(p)$ is a transverse polarization vector \cite{Brown:1983bc}.

\subsection{Properties of the Ritus matrices}

The mode solution is supported on-shell, since
\begin{equation}
    \label{eq: on-shell_support}
    \left[(\Pi_q^2 - M_W^2) \delta^{\mu}_{\, \nu} - 2 i q \mathcal{F}^{\mu}_{\, \nu}\right] w^\nu_q = (p^2 - M_W^2) w^\mu_q,
\end{equation}
because of the property of the Ritus matrix
\begin{equation}
    \label{eq: Ritus_diagonalization_EOM_operator}
    \left[\Pi_q^2 \delta^{\alpha}_{\, \beta} - 2 i q \mathcal{F}^{\alpha}_{\, \beta}\right] (E_q)^\beta_{\,\gamma} = p^2 (E_q)^\alpha_{\, \gamma},
\end{equation}
to diagonalize the EOM operator, proved in Appendix \ref{app: Ritus_properties}.

In addition to diagonalizing the equation of motion operator, the Ritus matrix satisfies an equation analogous to the fermionic case in \cite{Ritus:1985vta}, that reads
\begin{equation}
\begin{aligned}
    \label{eq: dressed_momentum_general}
    \Pi_q^\alpha(\phi) (E_q)^\mu{}_\nu(p,x)
    &=
    \pi_q^\alpha(p, \phi) (E_q)^\mu{}_\nu(p,x)\\
    &+
    i\,n^\alpha\frac{q}{n\cdot p}\,
    \mathcal F^\mu{}_\rho(\phi)\,
    (E_q)^\rho{}_\nu(p,x),
\end{aligned}
\end{equation}
such that, when contracting the momentum operator from Eq.~\eqref{def: momentum_operator} with the Ritus matrices and using $n^2 = 0$, 
\begin{equation} 
\label{eq: spectral_relations_Ritus}
\begin{aligned}
    \Pi_q^\mu (\phi) (E_q)_{\mu\nu} (p, x) &= \pi_q^\mu (p, \phi) (E_q)_{\mu\nu} (p, x),\\\Pi_q^\nu (\phi) (E_q)_{\mu\nu} (p, x) &= (E_q)_{\mu\nu} (p, x) \pi_q^\nu (p, \phi),
\end{aligned}
\end{equation}
where the dressed momentum for the charged vector boson is
\begin{equation}
    \label{eq: dressed_momentum_contracted}
    \pi^\mu_q (p, \phi) = p^\mu - q \mathcal{A}^\mu + n^\mu \frac{q (p \cdot \mathcal{A} (\phi))}{(n \cdot p)} - n^\mu \frac{\mathcal{A}^2}{2 (n \cdot p)},
\end{equation}
similar in form to the fermionic case described in \cite{Ritus:1985vta}. The relations shown in Eqs.~\eqref{eq: dressed_momentum_general}, \eqref{eq: spectral_relations_Ritus} are proved in Appendix \ref{app: Ritus_properties}.

For residual translations satisfying $\xi\cdot n=0$, 
\begin{equation}
\begin{aligned}
    [T(\xi)E_q(p,\cdot)] (x)
    &\equiv
    (E_q)^\mu{}_\nu (p,x+\xi)\\
    &=
    e^{-ip\cdot\xi}\,
    (E_q)^\mu{}_\nu (p,x),
\end{aligned}
\end{equation}
which yields infinitesimally, or directly contracting Eq.~\eqref{eq: dressed_momentum_general},
\begin{equation}
\begin{aligned}
    (\xi \cdot P) (E_q)_{\mu\nu} (p, x) &= (\xi \cdot p) (E_q)_{\mu\nu} (p, x),
\end{aligned}
\end{equation}
Thus, the Ritus matrices diagonalize the one-particle differential representation $\xi \cdot P$ of the residual-translation generator associated with the finite transformation $T(\xi)$, with eigenvalue $\xi\cdot p$. This is the one-particle representation of the same residual translation symmetry whose conserved Noether charge is $\mathcal Q_\xi = \xi\cdot P_{\mathrm{can}}$, as defined in Eq.~\eqref{eq: conserved_charge_phase_translation}. In particular, choosing $\xi = n$ the wave vector and $\xi = a_i$, $i = 1, 2$, where $a_i$ are the transverse basis vectors, leads to the spectral relations
\begin{equation}
\begin{aligned}
    P^+ (E_q)_{\mu\nu} (p, x) &= p^+ (E_q)_{\mu\nu} (p, x),\\
    P^{\perp, i} (E_q)_{\mu\nu} (p, x) &= p^{\perp, i} (E_q)_{\mu\nu} (p, x),
\end{aligned}
\end{equation}
where $X^+ = n \cdot X$ and $X^{\perp, i} = - a_i \cdot X$, for $X \in \{P, p\}$. 

As in the fermionic case from Ref.~\cite{Patuleanu:2021}, the perpendicular and phase-aligned (component of the) momentum of the charged boson is not affected by the Volkov background.




\subsection{Classical on-shell Volkov--Proca states}

The solution to the original Eq.~\eqref{eq: EOM} is
\begin{equation}
    W_q^\mu(x)
    =
    \int \frac{d^4p}{(2\pi)^4}\,
    (2\pi)\,
    \delta(p^2-M_W^2)\,
    c_q (p)\,
    w_q^\mu(p,x),
    \label{eq: Volkov_Proca_covariant_expansion}
\end{equation}
where the $\delta(p^2 - M^2_W)$ and the $(2\pi)$ factor can be inserted without loss of generality, making use of the support of the solution only on the mass shell $p^2 = M_W^2$ and that $c_q (p)$ is an arbitrary expansion coefficient. This solution is the (most) general solution, if (and only if) the mode functions $\{ w_q (p, x) \}_p$ form a complete set, as shown in the next section.

However, first, notice that using the property
\begin{equation}
    \delta (p^2 - M^2_W) = \frac{1}{2 \varepsilon_{\boldsymbol{p}}} \left[\delta(p_0 - \varepsilon_{\boldsymbol{p}}) + \delta(p_0 + \varepsilon_{\boldsymbol{p}})\right],
\end{equation}
where $\varepsilon_{\boldsymbol{p}} = \sqrt{\boldsymbol{p}^2 + M_W^2}$, the expansion from Eq.~\eqref{eq: Volkov_Proca_covariant_expansion}, after $p^0-$integration and a substitution $\boldsymbol{p} \to - \boldsymbol{p}$ for the negative frequency part, corresponding to $\delta (p_0 + \varepsilon_{\boldsymbol{p}})$, reads on-shell
\begin{equation}
    W_q^\mu(x)
    =
    \int
    \frac{d^3p}{(2\pi)^3\,2\varepsilon_{\boldsymbol p}}\,
    \sum_{r=\pm}
    c_q^r(\boldsymbol p)\,
    w_q^{r\,\mu}(\boldsymbol p,x),
    \label{eq: Volkov_Proca_on_shell_expansion}
\end{equation}
with a sum running over $r = \pm 1$ frequency sectors, and
\begin{equation} 
\begin{aligned}
w_q^{r\,\mu}(\boldsymbol p,x) 
&= w_q^\mu \!\left( p^0=r\varepsilon_{\boldsymbol p}, r\boldsymbol p;x \right) \\
&= E_q^\mu{}_\nu \!\left( r\varepsilon_{\boldsymbol p}, r\boldsymbol p;x \right) \varepsilon^\nu \!\left( r\varepsilon_{\boldsymbol p}, r\boldsymbol p \right),
\end{aligned}
\end{equation}
and a corresponding expression for the on-shell coefficients
\begin{equation} 
\label{eq: expansion_coeff_on_shell_classical}
c_q^r(\boldsymbol p) = c_q\!\left( p^0=r\varepsilon_{\boldsymbol p}, r\boldsymbol p \right).
\end{equation}

Complex conjugation relates the frequency sectors by
\begin{equation}
    \left[
        w_{q}^{r\,\mu}(\boldsymbol{p},x)
    \right]^*
    =
    w_{-q}^{-r\,\mu}(\boldsymbol{p},x).
\label{eq: Volkov_Proca_complex_conjugation}
\end{equation}

\section{Completeness and orthonormality}
\label{sec: completeness_orthonormality}

For a generic timelike spectral momentum $p$, three vectors $\{\varepsilon^\mu_{(\lambda)}(p)\}_{\lambda=1,2,3}$ may be chosen to span the subspace transverse to $p^\mu$, as described in \cite{GreinerReinhardt1996, Weinberg:1995}.

\subsection{Polarization completeness relation}
\label{subsec: completeness_dressed_polarizations}

The outer-product sum of the transverse polarization vectors defines the numerator tensor \cite{GreinerReinhardt1996},
\begin{equation}
\label{eq: completeness_free_polarizations}
   \mathscr{P}^{\mu\nu} (p; s) = \sum_{\lambda = 1}^3 \varepsilon^{\mu}_{(\lambda)} (p) \, \bar{\varepsilon}^{\nu}_{(\lambda)} (p) = - \eta^{\mu\nu} + \frac{p^\mu p^\nu}{s},
\end{equation}
with $s = p^2$. Off-shell, this is understood as a numerator tensor rather than as the completeness relation of physical Proca polarizations, while on the mass shell, $s = M_W^2$, it reduces to the usual polarization-completeness tensor of the free Proca field.

The mode solution in Eq.~\eqref{eq: EOM_mode_soln_off_shell} may
equivalently be interpreted in a passive form, whereby the free Proca
polarization vectors are transformed by the corresponding Lorentz
little-group element,
\begin{equation}
    \epsilon^\mu_{q,(\lambda)}(p,\phi)
    =
    \Lambda_q^\mu{}_{\nu}(p,\phi)\,
    \varepsilon^\nu_{(\lambda)}(p).
\end{equation}
The corresponding dressed numerator tensor is therefore
\begin{equation}
\label{eq: completeness_dressed_polarizations}
\begin{aligned}
    \mathscr P_q^{\mu\nu}(p,\phi; s)
    &\coloneqq
    \sum_{\lambda=1}^{3}
    \epsilon^\mu_{q,(\lambda)}(p,\phi)\,
    \bar\epsilon^\nu_{q,(\lambda)}(p,\phi)
    \\
    &=
    -\eta^{\mu\nu}
    +
    \frac{
        \pi_q^\mu(p,\phi)\,
        \pi_q^\nu(p,\phi)
    }{s}.
\end{aligned}
\end{equation}
with $s = p^2$, which, as before, represents on-shell the completeness relation for the three dressed polarizations. Hence, the Ritus matrices (or the little group elements) convert the free numerator tensor into the dressed one,
\begin{equation}
\label{eq: dressed_Proca_numerator_Ritus}
\begin{aligned}
    \mathscr P_q^{\mu\nu}(p,\phi_x; s)
    &=
    E_q^\mu{}_{\alpha}(p,x)\,
    \mathscr P^{\alpha\beta}(p)\,
    \bar E_{q,\beta}{}^\nu(p,x)
    \\
    &=
    \Lambda_q^\mu{}_{\alpha}(p, \phi_x)\,
    \mathscr P^{\alpha\beta}(p; s)\,
    \Lambda_{q,\beta}{}^\nu(p,\phi_x),
\end{aligned}
\end{equation}
where $\phi_x\equiv n\cdot x$. Consequently, for fixed on-shell spectral momentum $p$, the polarization completeness relation ($s = p^2 = M_W^2)$ for the Volkov--Proca modes is
\begin{equation}
\label{eq: completeness_dressed_Volkov_Proca_modes}
    \sum_{\lambda=1}^{3}
    w_{q, \lambda}^\mu(\boldsymbol{p},x)\,
    \bar w_{q, \lambda}^\nu(\boldsymbol{p},x)
    =
    \mathscr P_q^{\mu\nu}(\boldsymbol{p},\phi_x; M_W^2),
\end{equation}
where $w^\mu_{q, \lambda} (\boldsymbol{p}, x)$ is associated to the basis vector $\varepsilon^\nu_{(\lambda)} (\boldsymbol{p})$. Moreover, the normalization of the free polarization vectors is preserved by the Volkov dressing, that is
\begin{equation}
\begin{aligned}
    \bar w_{q, \lambda\,\mu}(\boldsymbol{p},x)\,
    w_{q, \lambda'}^\mu(\boldsymbol{p},x)
    &=
    \bar\varepsilon_{(\lambda)\mu}(\boldsymbol{p})\,
    \varepsilon_{(\lambda')}^\mu(\boldsymbol{p})
    \\
    &=
    -\delta_{\lambda\lambda'},
\end{aligned}
\label{eq: VP_mode_normalization_EMT}
\end{equation}
under the normalization of the free Proca polarizations.

\subsection{Orthogonality of the Volkov--Proca states}
\label{subsec: orthogonality_states}

The sesquilinear product for on-shell modes of same $q$ is
\begin{equation}
\begin{aligned}
\bigl(w_{r,\lambda}(\boldsymbol p),
      w_{r',\lambda'}(\boldsymbol p')\bigr)
&=
-i\int d^3x\,
\bar w_{r,\lambda,\mu}(\boldsymbol p,x)
\\[-1mm]
&\qquad
\overleftrightarrow{\partial}^{\,0}
w_{r',\lambda'}^{\mu}(\boldsymbol p',x),
\end{aligned}
\end{equation}
with the orthogonality relations as in the free case \cite{Cotaescu:2026}
\begin{equation}
\label{eq: orthogonality_relations_classical}
\begin{aligned}
    \bigl(
        w_{r,\lambda}(\boldsymbol p),
        w_{r',\lambda'}(\boldsymbol p')
    \bigr)
    &=
    r\,(2\pi)^3\,2\varepsilon_{\boldsymbol p}\,\\
    &\times\delta_{rr'}\delta_{\lambda\lambda'}
    \delta^{(3)}(\boldsymbol p-\boldsymbol p'),
\end{aligned}
\end{equation}
making use of the time-independence from Eq.~\eqref{eq: inner_product_time_independence}.

The mode coefficients are obtained by projecting the solution onto the
corresponding on-shell modes, using the orthogonality relations from Eq.~\eqref{eq: orthogonality_relations_classical} above
\begin{equation}
    \label{eq: mode_coefficients_innner_product}
    c_{q,\lambda}^r(\boldsymbol p)
    =
    r\,
    \bigl(
        w_{q,r,\lambda}(\boldsymbol p),
        W_q
    \bigr).
\end{equation}

The sesquilinear form is not positive definite: for two solutions
$W_q^{(1)}$ and $W_q^{(2)}$, expanded in the frequency sectors
$r=\pm1$, one finds
\begin{equation}
    \bigl(W_q^{(1)},W_q^{(2)}\bigr)
    =
    \sum_{r=\pm1}
    r\,
    \bigl(c_q^{(1),r},c_q^{(2),r}\bigr),
\end{equation}
where
\begin{equation}
\begin{aligned}
    \bigl(c_q^{(1),r},c_q^{(2),r}\bigr)
    &=
    \int
    \frac{d^3p}{(2\pi)^3\,2\varepsilon_{\boldsymbol p}}
    c_{q}^{(1),r*}(\boldsymbol p)\,
    c_{q}^{(2),r}(\boldsymbol p)\\
    &=
    \int
    \frac{d^3p}{(2\pi)^3\,2\varepsilon_{\boldsymbol p}}
    \sum_{\lambda=1}^3
    c_{q,\lambda}^{(1),r*}(\boldsymbol p)\,
    c_{q,\lambda}^{(2),r}(\boldsymbol p).
\end{aligned}
\end{equation}

In particular, as for the free complex vector boson \cite{Proca:1936fbw, Cotaescu:2026},
\begin{equation}
    \bigl(W_q,W_q\bigr)
    =
    \bigl(c_q^{+},c_q^{+}\bigr)
    -
    \bigl(c_q^{-},c_q^{-}\bigr),
\end{equation}
which makes manifest that the sesquilinear form is indefinite.

\subsection{Completeness and orthonormality of Ritus matrices}
\label{subsec: completeness_orthonormality_Ritus}

The Ritus matrices form a complete basis in configuration space. Their
completeness relations are
\begin{equation}
\label{eq: Ritus_completeness}
\begin{aligned}
    \int \frac{d^4 p}{(2\pi)^4} E^\mu_{\, \alpha} (p, x)\, \eta^{\alpha\beta}\, \bar{E}^\nu_{\, \beta} (p, x') &= \eta^{\mu\nu} \delta^4 (x - x'),\\
    \int \frac{d^4 p}{(2\pi)^4} \bar{E}^\mu_{\, \alpha} (p, x)\, \eta^{\alpha\beta}\, E^\nu_{\, \beta} (p, x') &= \eta^{\mu\nu} \delta^4 (x - x'),
\end{aligned}
\end{equation}
between Ritus matrices, using $\bar{\cdot}$ to denote complex conjugation of the underlying variable in both expressions above.

The corresponding orthonormality relations are
\begin{equation}
\begin{aligned}
    \int d^4 x\, \bar{E}^\mu_{\, \alpha} (p, x)\, \eta^{\alpha\beta}\, E^\nu_{\, \beta} (p', x) &= \eta^{\mu\nu} (2 \pi)^4 \delta^4 (p - p'),\\
    \int d^4 x\, E^\mu_{\, \alpha} (p, x)\, \eta^{\alpha\beta}\, \bar{E}^\nu_{\, \beta} (p', x) &= \eta^{\mu\nu} (2 \pi)^4 \delta^4 (p - p').
\end{aligned}
\label{eq: Ritus_orthogonality}
\end{equation}
Thus, while Eq.~\eqref{eq: Ritus_completeness} provides a resolution of
the identity in configuration space, Eq.~\eqref{eq: Ritus_orthogonality}
expresses orthonormality with respect to the spectral-momentum labels.

These relations are proved in Appendix \ref{app: completeness_ortho_Ritus}, by integrating in light-cone coordinates over momenta for the completeness relations and over configurations for orthogonality, respectively.

\section{Volkov--Proca propagator}
\label{sec: Volkov-Proca_propagator}

The Green equation for the Volkov--Proca theory reads
\begin{equation}
    \label{eq: Green}
    (\mathcal{O}_q)^{\alpha}{}_{\beta}\, G^\beta_{q\, \gamma} (x, x^\prime) = - \delta^\alpha_{\, \gamma} \delta^{(4)} (x - x^\prime),
\end{equation}
under the transversality constraint given by
\begin{equation}
    \label{eq: transversality_constraint_Green_function}
    (\Pi_q)_\alpha G^\alpha_{\, \gamma} (x, x^\prime) = \frac{1}{M^2_W} (\Pi_q)_\alpha \left[\delta^\alpha_{\, \gamma} \delta^{(4)} (x - x^\prime)\right],
\end{equation}

This is obtained by adding\footnote{The source terms corresponding to Eq.~\eqref{eq: Green} are actually subtracted.} a source term $J_q$ (for each charge sector) in the Volkov-Proca Lagrangian, leading to an Euler-Lagrange equation
\begin{equation}
    (\mathcal{O}_q)^{\alpha}{}_{\beta} W_q^\beta (x) = - J_q^\alpha (x),
\end{equation}
under the transversality constraint obtained as in Eq.~\eqref{eq: transversality_constraint},
\begin{equation}
    \label{eq: transversality_condition_with_source}
    (\Pi_q)_\alpha W_q^\alpha (x) = \frac{1}{M^2_W} (\Pi_q)_\alpha J_q^\alpha (x),
\end{equation}
and expanding out the solution,
\begin{equation}
    W^\beta_q (x) = \int d^4 x^\prime\, G^\beta_{q\, \gamma} (x, x^\prime)\, J_q^\gamma (x^\prime).
\end{equation}

In deriving the transversality constraint from Eqs.~\eqref{eq: transversality_condition_with_source} and \eqref{eq: transversality_constraint_Green_function}, the term involving the background vanishes since for a plane-wave configuration $\partial_\alpha \mathcal F^{\alpha}{}_\beta (\phi) = n_\alpha \mathcal F^{\prime\alpha}{}_\beta (\phi) = 0$.

\subsection{Covariant Volkov--Proca propagator}

Plugging in Green equation from Eq.~\eqref{eq: Green} the ansatz
\begin{equation}
    \label{eq: covariant_propagator}
    G^{\mu\nu} (x, x^\prime) = \int \frac{d^4 p}{(2 \pi)^4} E^\mu_{\,\rho} (p, x)\, G^{\rho\sigma} (p)\, \bar{E}^\nu_{\, \sigma} (p, x'),
\end{equation}
and using the Ritus diagonalization property from Eq.~\eqref{eq: Ritus_diagonalization_EOM_operator} on the left side and the completeness of the Ritus matrices from Eq.~\eqref{eq: Ritus_completeness} to write a resolution of the identity on the right hand side, together with the relation for the longitudinal term
\begin{equation}
\begin{aligned}
    \Pi^\alpha \Pi_\beta\, G^{\beta\gamma} (x, x^\prime) 
    &= \int \frac{d^4 p}{(2\pi)^4} E^\alpha_{\,\rho} (p, x) \frac{p^\rho p^\sigma}{M_W^2} \bar{E}^\gamma_{\, \sigma} (p, x'),
\end{aligned}
\end{equation}
as proved in Eq.~\eqref{eq: Green_longitudinal_second_derivative}, leads to the solution kernel
\begin{equation}
    G^{\rho\sigma} (p) \sim \frac{- \eta^{\rho\sigma} + p^\rho p^\sigma / M_W^2}{p^2 - M^2_W},
\end{equation}
where the symbol $\sim$ indicates that the Green function is specified only after imposing the boundary condition, corresponding to a choice of how the poles at $p^0=\pm\varepsilon_{\boldsymbol p}$ are enclosed. Equivalently, one may retain the $p^0$ integration along the real axis and implement the appropriate pole-shifting prescription,
\begin{equation}
\begin{aligned}
    G_{A/R}^{\rho\sigma}(p)
    &=
    \frac{-\eta^{\rho\sigma}+p^\rho p^\sigma/M_W^2}
    {(p^0\mp i0)^2-\varepsilon_{\boldsymbol p}^2},
    \\
    G_F^{\rho\sigma}(p)
    &=
    \frac{-\eta^{\rho\sigma}+p^\rho p^\sigma/M_W^2}
    {p^2-M_W^2+i0},
\end{aligned}
\label{eq: Volkov_Proca_momentum_space_Green_kernels}
\end{equation}
where the upper and lower signs correspond to the advanced and retarded prescriptions, respectively, while the Feynman Green function is defined by the $(+i 0)$ prescription.

\subsubsection*{Representation in terms of the scalar Volkov propagator}

To express the Volkov--Proca propagator in terms of the scalar Volkov propagator, we first introduce
\begin{equation}
    \label{eq: scalar_Ritus_factor}
    \mathscr E_q(p,x)
    \coloneqq
    e^{-ip\cdot x}U_q(p,\phi),
\end{equation}
the Ritus factor containing the scalar Volkov dressing. The corresponding momentum-space kernel of the scalar Volkov propagator \cite{Berestetskii:1982qgu, Brown:1983bc} is defined as
\begin{equation}
    \mathcal D_{F,q}(p;x,x')
    \coloneqq
    \frac{
        \mathscr E_q(p,x)
        \bar{\mathscr E}_q(p,x')
    }{
        p^2-M_W^2+i0
    },
    \label{eq: scalar_Volkov_kernel}
\end{equation}
so that the scalar Volkov propagator takes the form
\begin{equation}
    \Delta_{F,q}(x,x')
    =
    \int\frac{d^4p}{(2\pi)^4}\,
    \mathcal D_{F,q}(p;x,x').
    \label{eq: scalar_Volkov_propagator}
\end{equation}

Using the factorization of the vector Ritus matrices into their scalar and Lorentz parts, the Volkov--Proca propagator may then be written as
\begin{equation}
\begin{aligned}
    G_{F,q}^{\mu\nu}(x,x')
    ={}&
    \int\frac{d^4p}{(2\pi)^4}\,
    \mathcal D_{F,q}(p;x,x')\,
    \mathscr P_q^{\mu\nu}
    \bigl(p;\phi_x,\phi_{x'};M_W^2\bigr),
\end{aligned}
\end{equation}
where the bilocal Proca numerator is
\begin{equation}
\begin{aligned}
    \mathscr P_q^{\mu\nu}
    \bigl(p;\phi_x,\phi_{x'};M_W^2\bigr)
    \coloneqq{}&
    -
    \Lambda_q^{\mu\nu}
    \bigl(p;\phi_x,\phi_{x'}\bigr)
    \\
    &+
    \frac{
        \pi_q^\mu(p,\phi_x)\,
        \pi_q^\nu(p,\phi_{x'})
    }{
        M_W^2
    },
\end{aligned}
\label{eq: bilocal_dressed_Proca_numerator}
\end{equation}
in terms of the relative Lorentz transformation in App.~\ref{app: phase_transport_dressed_momentum}.

The Volkov--Proca propagator cannot, in general, be obtained from the scalar Volkov propagator through the naive replacement
\begin{equation}
G_{F,q}^{\mu\nu}(x,x')
\neq
\left(
-\eta^{\mu\nu}
+
\frac{\Pi_{q,x}^{\mu}\Pi_{q,x}^{\nu}}{M_W^2}
\right)
\Delta_{F,q}(x,x').
\end{equation}
Instead, it is obtained from the scalar Volkov propagator as
\begin{equation}
\begin{aligned}
G_{F,q}^{\mu\nu}(x,x')
={}&
-
\Lambda_q^{\mu\nu} \bigl(P_x; \phi_x,\phi_{x'}\bigr)\,
\Delta_{F,q}(x,x')
\\
&+
\frac{1}{M_W^2}\,
\Pi_{q,x}^\mu\,
\Delta_{F,q}(x,x')\,
\overleftarrow{\Pi}_{q,x'}^\nu,
\end{aligned}
\label{eq: Volkov_Proca_scalar_Volkov_representation}
\end{equation}
where $\Lambda_q^{\mu\nu}\bigl(P_x;\phi_x,\phi_{x'}\bigr)$
denotes the operator-valued counterpart of $\Lambda_q^{\mu\nu}\bigl(p;\phi_x,\phi_{x'}\bigr)$, obtained through the spectral replacement $n\cdot p\to n\cdot P_x$.

\subsection{On-shell Volkov--Proca propagator}

The $p^0$-dependent part of the integrand in Eq.~\eqref{eq: covariant_propagator},
\begin{equation}
    f^{\mu\nu}(p^0)
    =
    \frac{
        e^{-ip^0(t-t')}
    }{
        (p^0)^2-\varepsilon_{\boldsymbol p}^{\,2}
    }\,
    \mathcal N^{\mu\nu}(p;\phi,\phi'),
\end{equation}
where the Ritus-dressed Proca numerator given by
\begin{equation}
    \label{eq: phase_kernel_p}
    \mathcal N^{\mu\nu}(p;\phi,\phi')
    \coloneqq
    E^\mu{}_{\rho}(p,\phi)\,
    \mathscr P^{\rho\sigma}(p; M_W^2)\,
    \bar E^\nu{}_{\sigma}(p,\phi'),
\end{equation}
exhibits the two usual mass-shell poles at $p^0=\pm\varepsilon_{\boldsymbol p}$, with $\varepsilon_{\boldsymbol p}=\sqrt{\boldsymbol p^2+M_W^2}$, together with an essential singularity at $p^0=\boldsymbol n\cdot\boldsymbol p$, or equivalently at $n\cdot p=0$, originating from the Ritus dressing. Nevertheless, the $p^0$ integration can be performed (only for the decaying term, as explained next), deforming the contour around the singularity of the Ritus dressing at $n\cdot p=0$ so that the singularity lies outside the closed contour. This prescription is consistent with the physical massive spectrum, since the mass-shell poles satisfy \(n\cdot p\neq0\) and therefore do not coincide with the Volkov singularity. With \(n\cdot p=0\) excluded from the closed contour by the prescribed deformation, the only enclosed singularities are the relevant mass-shell poles, namely
\begin{align}
    \underset{p^0=\varepsilon_{\boldsymbol p}}
    {\operatorname{Res}}\,
    f^{\mu\nu}(p^0)
    &=
    \frac{
        e^{-i\varepsilon_{\boldsymbol p}(t-t')}
    }{
        2\varepsilon_{\boldsymbol p}
    }\,
    \mathcal N^{\mu\nu}
    (\varepsilon_{\boldsymbol p},\boldsymbol p;\phi,\phi'),
    \\
    \underset{p^0=-\varepsilon_{\boldsymbol p}}
    {\operatorname{Res}}\,
    f^{\mu\nu}(p^0)
    &=
    -
    \frac{
        e^{+i\varepsilon_{\boldsymbol p}(t-t')}
    }{
        2\varepsilon_{\boldsymbol p}
    }\,
    \mathcal N^{\mu\nu}
    (-\varepsilon_{\boldsymbol p},\boldsymbol p;\phi,\phi').
\end{align}

The application of the residue theorem requires some care, since the longitudinal part of the Proca numerator contains a term that does not vanish as $|p^0|\to\infty$. We therefore separate the integrand, denoting $    \tau\coloneqq t-t'$, as
\begin{equation}
    f^{\mu\nu}(p^0)
    =
    f^{\mu\nu}_{\mathrm{ct}} (p^0) 
    +
    f_{\mathrm{dec}}^{\mu\nu}(p^0),
\end{equation}
where the contribution that generates the contact term
\begin{equation}
    f^{\mu\nu}_{\mathrm{ct}} (p^0) \coloneqq
    \frac{\mathcal U_q(p;\phi,\phi')}{M_W^2}\,
    \delta^\mu{}_0\delta^\nu{}_0\,
    e^{-ip^0\tau},
\end{equation}
with the bilocal Volkov phase factor,
\begin{equation}
\mathcal U_q (p;\phi,\phi')
\coloneqq
U_q(p,\phi)\,
\bar U_q(p,\phi'),
\end{equation}
does not vanish as $|p^0|\to\infty$, whereas $f_{\mathrm{dec}}^{\mu\nu}(p^0)$ decays with increasing $|p^0|$ and, for $\tau\neq0$, its contribution from the large semicircle vanishes.

The exponential factor inside $f_{\mathrm{dec}}^{\mu\nu}(p^0)$ fixes the contour closure: for $\tau>0$ the contour is closed in the lower half-plane, whereas for $\tau<0$ it is closed in the upper half-plane. Accordingly, after
implementing the prescribed deformation around the Volkov singularity $n\cdot p=0$, or, equivalently, shifting the singularity and integrating along the real line, the decaying part gives
\begin{equation}
    \int_\mathbb{R}
    \frac{dp^0}{2\pi}\,
    f_{\mathrm{dec}}^{\mu\nu}(p^0)
    =
    \begin{cases}
        -i\displaystyle\sum_{\mathrm{LHP}}
        \operatorname{Res} f_{\mathrm{dec}}^{\mu\nu},
        & \tau>0,
        \\[2mm]
        +i\displaystyle\sum_{\mathrm{UHP}}
        \operatorname{Res} f_{\mathrm{dec}}^{\mu\nu},
        & \tau<0.
    \end{cases}
\end{equation}

The remaining contribution $f_{\mathrm{ct}}^{\mu\nu}(p^0)$ is not conveniently evaluated by a $p^0$-contour argument, because the bilocal Volkov factor
$\mathcal U_q(p;\phi,\phi')$ retains the singularity at $n\cdot p=0$.
Instead, it can be evaluated directly at the level of the complete
four-momentum integral. Introducing
$\Delta x^\mu=x^\mu-x^{\prime\mu}$, one has
\begin{equation}
\begin{aligned}
    I_{\mathrm{ct}}^{\mu\nu}(x,x')
    \equiv{}&
    \frac{\delta^\mu{}_0\delta^\nu{}_0}{M_W^2}
    \int\frac{d^4p}{(2\pi)^4}\,
    e^{-ip\cdot\Delta x}\,
    \mathcal U_q(p;\phi,\phi').
\end{aligned}
\end{equation}

In the gauge $\mathcal A^+=0$, the Volkov factor is independent of $p^-$. Hence, writing $d^4p = dp^+\,dp^-\,d^2\boldsymbol p^\perp$, the $p^-$ integration therefore gives
\begin{equation}
    \int\frac{dp^-}{2\pi}\,
    e^{-ip^-(x^+-x^{\prime +})}
    =
    \delta(x^+-x^{\prime +}).
\end{equation}
On the support of this distribution the phases coincide,
\begin{equation}
    \mathcal U_q(p;\phi,\phi')
    \big|_{\phi=\phi'}
    =
    1.
\end{equation}
The remaining Fourier integrations consequently yield
\begin{equation}
\begin{aligned}
    I_{\mathrm{ct}}^{\mu\nu}(x,x')
    &=
    \frac{\delta^\mu{}_0\delta^\nu{}_0}{M_W^2}\,
    \delta^{(4)}(x-x'),
\end{aligned}
\label{eq: dressed_Proca_contact_term}
\end{equation}
which does not correspond to a propagating physical degree of freedom, but is a local contact term required for the propagator to satisfy the Green equation.

Consequently, although the nondecaying part retains the bilocal Volkov dressing before the momentum integrations are performed, its complete four-momentum contribution reduces to the same local contact term as in the free Proca theory.

For the retarded prescription, both poles lie in the lower half-plane.
Therefore, for $\tau>0$ the contour is closed clockwise around both
poles, whereas for $\tau<0$ no pole is enclosed, such that 

\begin{widetext}
\begin{equation}
\begin{aligned}
    \int_{\mathbb R}\frac{dp^0}{2\pi}\,
    f_{R, \mathrm{dec}}^{\mu\nu}(p^0)
    &=
    -i\theta(\tau)
    \left[
        \underset{p^0=\varepsilon_{\boldsymbol p}}
        {\operatorname{Res}}\,f^{\mu\nu}
        +
        \underset{p^0=-\varepsilon_{\boldsymbol p}}
        {\operatorname{Res}}\,f^{\mu\nu}
    \right] =
    -\frac{i\theta(\tau)}
    {2\varepsilon_{\boldsymbol p}}
    \Big[
        e^{-i\varepsilon_{\boldsymbol p}\tau}
        \mathcal N^{\mu\nu}
        (\varepsilon_{\boldsymbol p},\boldsymbol p;\phi,\phi')
        -
        e^{+i\varepsilon_{\boldsymbol p}\tau}
        \mathcal N^{\mu\nu}
        (-\varepsilon_{\boldsymbol p},\boldsymbol p;\phi,\phi')
    \Big].
\end{aligned}
\label{eq:retarded_p0_Cauchy}
\end{equation}

For the advanced prescription, both poles lie in the upper half-plane. Accordingly,
\begin{equation}
\begin{aligned}
    \int_{\mathbb R}\frac{dp^0}{2\pi}\,
    f_{A, \mathrm{dec}}^{\mu\nu}(p^0)
    &=
    i\theta(-\tau)
    \left[
        \underset{p^0=\varepsilon_{\boldsymbol p}}
        {\operatorname{Res}}\,f^{\mu\nu}
        +
        \underset{p^0=-\varepsilon_{\boldsymbol p}}
        {\operatorname{Res}}\,f^{\mu\nu}
    \right] =
    \frac{i\theta(-\tau)}
    {2\varepsilon_{\boldsymbol p}}
    \Big[
        e^{-i\varepsilon_{\boldsymbol p}\tau}
        \mathcal N^{\mu\nu}
        (\varepsilon_{\boldsymbol p},\boldsymbol p;\phi,\phi')
        -
        e^{+i\varepsilon_{\boldsymbol p}\tau}
        \mathcal N^{\mu\nu}
        (-\varepsilon_{\boldsymbol p},\boldsymbol p;\phi,\phi')
    \Big].
\end{aligned}
\label{eq:advanced_p0_Cauchy}
\end{equation}

For the Feynman prescription, the positive-energy pole is displaced into the lower half-plane, while the negative-energy pole is displaced into the upper half-plane, as shown in Figure~\ref{fig: feynman_p0_contours}. It follows that
\begin{equation}
\begin{aligned}
    \int_{\mathbb R}\frac{dp^0}{2\pi}\,
    f_{F, \mathrm{dec}}^{\mu\nu}(p^0)
    ={}&
    -i\theta(\tau)
    \underset{p^0=\varepsilon_{\boldsymbol p}}
    {\operatorname{Res}}\,f^{\mu\nu}
    +
    i\theta(-\tau)
    \underset{p^0=-\varepsilon_{\boldsymbol p}}
    {\operatorname{Res}}\,f^{\mu\nu}
    \\
    ={}&
    -\frac{i}{2\varepsilon_{\boldsymbol p}}
    \Big[
        \theta(\tau)
        e^{-i\varepsilon_{\boldsymbol p}\tau}
        \mathcal N^{\mu\nu}
        (\varepsilon_{\boldsymbol p},\boldsymbol p;\phi,\phi')
        +
        \theta(-\tau)
        e^{+i\varepsilon_{\boldsymbol p}\tau}
        \mathcal N^{\mu\nu}
        (-\varepsilon_{\boldsymbol p},\boldsymbol p;\phi,\phi')
    \Big].
\end{aligned}
\label{eq:Feynman_p0_Cauchy}
\end{equation}
\end{widetext}

Performing the change of variables $\boldsymbol p\to-\boldsymbol p$ in the negative-energy contribution, and using $\varepsilon_{-\boldsymbol p}=\varepsilon_{\boldsymbol p}$, the four-momentum associated with the negative-energy pole transforms as
\begin{equation*}
    \left(
        -\varepsilon_{\boldsymbol p},
        \boldsymbol p
    \right)
    \xrightarrow{\boldsymbol p\to-\boldsymbol p}
    \left(
        -\varepsilon_{\boldsymbol p},
        -\boldsymbol p
    \right)
    =
    -p^\mu,
\end{equation*}
such that the second term is then mapped as
\begin{equation*}
\begin{aligned}
    &e^{+i\varepsilon_{\boldsymbol p}(t-t')}
    e^{+i\boldsymbol p\cdot(\boldsymbol x-\boldsymbol x')}
    \mathcal N^{\mu\nu}
    (-\varepsilon_{\boldsymbol p},\boldsymbol p;\phi,\phi')
    \\
    &\qquad\longmapsto
    e^{+ip\cdot(x-x')}
    \mathcal N^{\mu\nu}
    (-p;\phi,\phi').
\end{aligned}
\end{equation*}
Since the polarization tensor is even under $p^\mu\to-p^\mu$,
\begin{equation}
    \mathscr P^{\rho\sigma}(-p)
    =
    \mathscr P^{\rho\sigma}(p),
\end{equation}
the two frequency sectors can be encoded uniformly as
\begin{equation}
\label{eq: phase_kernel_frequency_sector}
\begin{aligned}
    \mathcal N_r^{\mu\nu}
    (\boldsymbol p;\phi,\phi')
    &\coloneqq
    \mathcal N^{\mu\nu}
    (rp;\phi,\phi')
    \\
    &=
    E^\mu{}_{\rho}(rp,\phi)\,
    \mathscr P^{\rho\sigma}(p)\,
    \bar E^\nu{}_{\sigma}(rp,\phi'),
\end{aligned}
\end{equation}
with $r=\pm1$, where $p^\mu = \left(\varepsilon_{\boldsymbol p}, \boldsymbol p\right)$ and $\varepsilon_{\boldsymbol p} = \varepsilon_{-\boldsymbol p}$.

Consequently, the retarded and advanced propagators read
\begin{widetext}
\begin{equation}
\begin{aligned}
    iG_{R, \mathrm{dec}}^{\mu\nu}(x,x')
    &=
    \theta(t-t')
    \int d\Gamma_p\,
    \left[
        e^{-ip\cdot(x-x')}
        \mathcal N_+^{\mu\nu}
        (\boldsymbol p;\phi,\phi')
        -
        e^{+ip\cdot(x-x')}
        \mathcal N_-^{\mu\nu}
        (\boldsymbol p;\phi,\phi')
    \right],
    \\
    iG_{A, \mathrm{dec}}^{\mu\nu}(x,x')
    &=
    -\theta(t'-t)
    \int d\Gamma_p\,
    \left[
        e^{-ip\cdot(x-x')}
        \mathcal N_+^{\mu\nu}
        (\boldsymbol p;\phi,\phi')
        -
        e^{+ip\cdot(x-x')}
        \mathcal N_-^{\mu\nu}
        (\boldsymbol p;\phi,\phi')
    \right],
\end{aligned}
\end{equation}
with the (on-shell) Lorentz-invariant measure denoted by $d\Gamma_p \coloneqq d^3 \boldsymbol{p} / [{(2\pi)^3 \,2\varepsilon_{\boldsymbol p}}]$.

\begin{figure*}[t]
\centering

\begin{minipage}{0.48\textwidth}
\centering
\begin{tikzpicture}[x=1cm,y=1cm,>=Latex,baseline=(current bounding box.center)]
    \def\R{2.8}
    \def\eps{1.45}
    \def\np{0.85}
    \def\rho{0.24}

    \draw[->] (-3.35,0) -- (3.35,0) node[right] {$\Re\,p^0$};
    \draw[->] (0,-3.15) -- (0,3.15) node[above] {$\Im\,p^0$};

    \node[below left] at (0,0) {$0$};

    \draw (-\eps,0.06) -- (-\eps,-0.06);
    \draw (\eps,0.06) -- (\eps,-0.06);
    \draw (\np,0.06) -- (\np,-0.06);

    \node[below] at (-\eps,0) {$-\varepsilon_{\boldsymbol p}$};
    \node[above] at (\eps,0) {$\varepsilon_{\boldsymbol p}$};
    \node[above] at (\np,0) {$\boldsymbol n\!\cdot\!\boldsymbol p$};

    \fill (-\eps,0.18) circle (1.2pt);
    \fill (\eps,-0.18) circle (1.2pt);

    \node[above left] at (-\eps,0.18) {$-\varepsilon_{\boldsymbol p}+i0$};
    \node[below right] at (\eps,-0.18) {$\varepsilon_{\boldsymbol p}-i0$};

    \draw[line width=0.5pt] (\np-0.08,0.08) -- (\np+0.08,-0.08);
    \draw[line width=0.5pt] (\np-0.08,-0.08) -- (\np+0.08,0.08);

    \draw[line width=0.8pt] (-\R,0) -- (\np-\rho,0);
    \draw[line width=0.8pt] (\np-\rho,0) arc[start angle=180,end angle=360,radius=\rho];
    \draw[line width=0.8pt] (\np+\rho,0) -- (\R,0);

    \draw[->,line width=0.8pt] (-2.35,0) -- (-1.85,0);
    \draw[->,line width=0.8pt] (2.10,0) -- (2.60,0);

    \draw[line width=0.8pt] (\R,0) arc[start angle=0,end angle=-180,radius=\R];
    \draw[<-,line width=0.8pt] (0.65,-\R+0.08) arc[start angle=-77,end angle=-58,radius=\R];

    \node at (0,-3.55) {(a) $\tau=t-t'>0$};
\end{tikzpicture}
\end{minipage}
\hfill
\begin{minipage}{0.48\textwidth}
\centering
\begin{tikzpicture}[x=1cm,y=1cm,>=Latex,baseline=(current bounding box.center)]
    \def\R{2.8}
    \def\eps{1.45}
    \def\np{0.85}
    \def\rho{0.24}

    \draw[->] (-3.35,0) -- (3.35,0) node[right] {$\Re\,p^0$};
    \draw[->] (0,-3.15) -- (0,3.15) node[above] {$\Im\,p^0$};

    \node[below left] at (0,0) {$0$};

    \draw (-\eps,0.06) -- (-\eps,-0.06);
    \draw (\eps,0.06) -- (\eps,-0.06);
    \draw (\np,0.06) -- (\np,-0.06);

    \node[below] at (-\eps,0) {$-\varepsilon_{\boldsymbol p}$};
    \node[above] at (\eps,0) {$\varepsilon_{\boldsymbol p}$};
    \node[below] at (\np,0) {$\boldsymbol n\!\cdot\!\boldsymbol p$};

    \fill (-\eps,0.18) circle (1.2pt);
    \fill (\eps,-0.18) circle (1.2pt);

    \node[above left] at (-\eps,0.18) {$-\varepsilon_{\boldsymbol p}+i0$};
    \node[below right] at (\eps,-0.18) {$\varepsilon_{\boldsymbol p}-i0$};

    \draw[line width=0.5pt] (\np-0.08,0.08) -- (\np+0.08,-0.08);
    \draw[line width=0.5pt] (\np-0.08,-0.08) -- (\np+0.08,0.08);

    \draw[line width=0.8pt] (-\R,0) -- (\np-\rho,0);
    \draw[line width=0.8pt] (\np-\rho,0) arc[start angle=180,end angle=0,radius=\rho];
    \draw[line width=0.8pt] (\np+\rho,0) -- (\R,0);

    \draw[->,line width=0.8pt] (-2.35,0) -- (-1.85,0);
    \draw[->,line width=0.8pt] (2.10,0) -- (2.60,0);

    \draw[line width=0.8pt] (\R,0) arc[start angle=0,end angle=180,radius=\R];
    \draw[->,line width=0.8pt] (-0.65,\R-0.08) arc[start angle=103,end angle=122,radius=\R];

    \node at (0,-3.55) {(b) $\tau=t-t'<0$};
\end{tikzpicture}
\end{minipage}

\caption{Complex \(p^0\)-plane contours for the Feynman Volkov--Proca propagator. The mass-shell poles lie at \(p^0=\pm\varepsilon_{\boldsymbol p}\mp i0\), and the Ritus dressing introduces an essential singularity at \(p^0=\boldsymbol n\cdot\boldsymbol p\). In panel (a), corresponding to \(\tau>0\), the contour is closed in the lower half-plane and the real-axis contour is deformed below the essential singularity so that it is not enclosed. In panel (b), corresponding to \(\tau<0\), the contour is closed in the upper half-plane and the contour is deformed above the essential singularity.}
\label{fig: feynman_p0_contours}
\end{figure*}
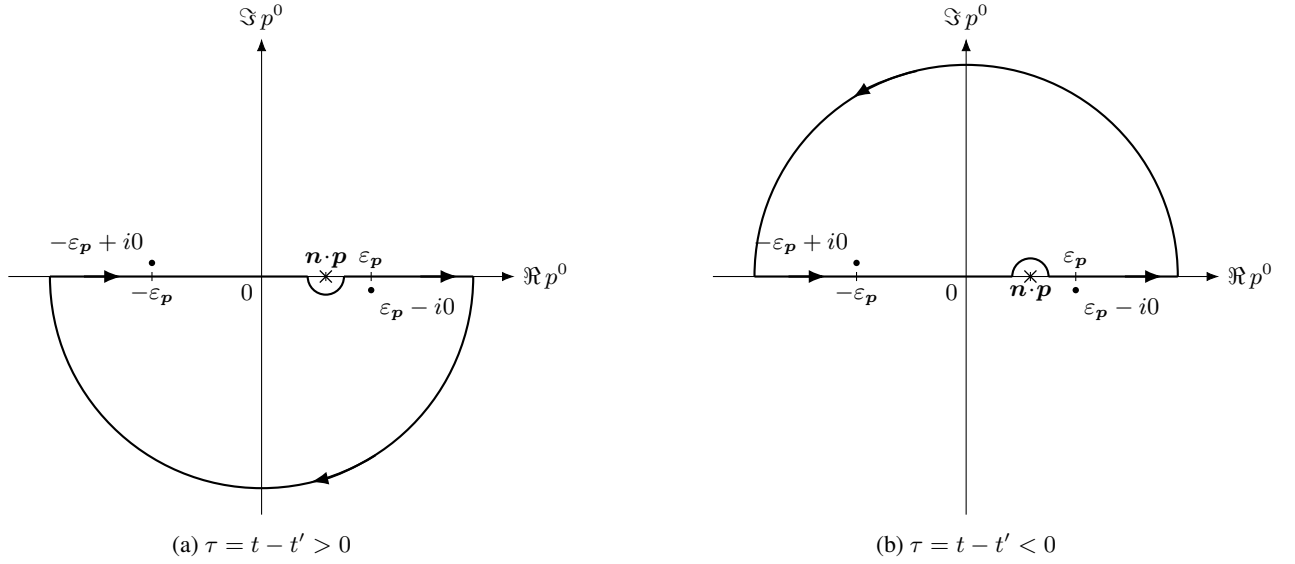

For the Feynman prescription, one obtains
\begin{equation}
\begin{aligned}
    iG_{F, \mathrm{dec}}^{\mu\nu}(x,x')
    ={}&
    \theta(t-t')
    \int d\Gamma_p\,
    e^{-ip\cdot(x-x')}
    \mathcal N_+^{\mu\nu}
    (\boldsymbol p;\phi,\phi') +
    \theta(t'-t)
    \int d\Gamma_p\,
    e^{+ip\cdot(x-x')}
    \mathcal N_-^{\mu\nu}
    (\boldsymbol p;\phi,\phi').
\end{aligned}
\end{equation}

\end{widetext}

The retarded propagator can be written in terms of the positive- and
negative-frequency Volkov--Proca modes as
\begin{equation}
    \label{eq: retarded_propagator_on_shell}
    G_{R, \mathrm{dec}}^{\beta}{}_{\gamma}(x,x')
    =
    -i\theta(t-t')\,
    \mathcal C_q^{\beta}{}_{\gamma}(x,x'),
\end{equation}
where the Pauli-Jordan function is
\begin{equation}
\begin{aligned}
    \mathcal C_q^{\beta}{}_{\gamma}(x,x')
    &\coloneqq
    \sum_{\lambda=1}^{3}
    \int d\Gamma_p
    \sum_{r=\pm1}r\,
    w_{q,r,\lambda}^{\beta}(\boldsymbol p,x)\,
    \bar w_{q,r,\lambda,\gamma}(\boldsymbol p,x').
\end{aligned}
\label{eq:Pauli_Jordan_mode_kernel_completeness}
\end{equation}

For $t\neq t'$ the full propagator is entirely determined
by the mass-shell pole contribution. We may therefore omit the subscript ``$\mathrm{dec}$'' in the following expressions describing propagation away from coincidence.

\section{Completeness of the Volkov--Proca states}
\label{sec: completenes_orthonormality_states}

The completeness relation can be obtained by acting with the Volkov--Proca operator from Eq.~\eqref{def: VP_EOM_operator} on the retarded propagator in Eq.~\eqref{eq: retarded_propagator_on_shell}
and restricting the Green equation to its purely spatial components,
$\alpha=i$ and $\gamma=j$,
\begin{equation}
    \mathcal{O}_{q,x}^{i}{}_{\beta}\,
    G_{R,q}^{\beta}{}_{j}(x,x')
    =
    -
    \delta^{i}{}_{j}\,
    \delta^{(4)}(x-x'),
\end{equation}
where the subscript $x$ indicates explicitly that the differential operator acts on the first argument. Since the contact contribution from Eq.~\eqref{eq: dressed_Proca_contact_term} is purely temporal in its second tensor index, one has
\begin{equation}
    G_{R,\mathrm{ct},q}^{\beta}{}_{j}(x,x')
    =
    0,
    \qquad
    j=1,2,3,
\end{equation}
such that, in the purely spatial sector the full retarded propagator
coincides with its propagating part,
\begin{equation}
    G_{R,q}^{\beta}{}_{j}(x,x')
    =
    -i\theta(t-t')\,
    \mathcal C_q^{\beta}{}_{j}(x,x').
\end{equation}
This equality holds distributionally, including at $t=t'$, since the
contact contribution vanishes identically for a spatial second index.

The term in which the Green operator passes through the step function of Eq.~\eqref{eq: retarded_propagator_on_shell} vanishes, that is
\begin{equation}
    -i\theta(t - t')\,
    \mathcal O_{q,x}^{i}{}_{\beta}\,
    \mathcal C_q^{\beta}{}_{j}(x,x') = 0,
\end{equation}
as the Volkov-Proca modes satisfy the homogeneous equation,
\begin{equation}
\label{eq:SolutionEOM}
    \mathcal O_{q,x}^{\alpha}{}_{\beta}\,
    w_{q,r,\lambda}^{\beta}(\boldsymbol{p}, x)
    =
    0,
\end{equation}
leaving only ``singular" terms acting on the step function. In simplified notation, the ``singular" kinetic term reads
\begin{align}
    \label{eq: kinetic_singular}
    \left[
        \Pi_q^2
        G_{R,q}^{i}{}_{j}
    \right]_{\mathrm{sing}}
    =
    i\delta'(\tau)
    \mathcal C_q^{i}{}_{j}
    +
    2i\delta(\tau)
    \partial_t
    \mathcal C_q^{i}{}_{j},
\end{align}
with $\tau = t - t'$, and the longitudinal, similarly, reads
\begin{equation}
    \left[
        -
        \Pi_q^i\Pi_{q,\beta}
        G_{R,q}^{\beta}{}_{j}
    \right]_{\mathrm{sing}}
    =
    -
    \delta(\tau)
    \Pi_q^i
    \mathcal C_q^{0}{}_{j},
    \label{eq: longitudinal_singular}
\end{equation}
such that, collecting all above, leads to
\begin{equation}
    \label{eq: singular_terms_collected}
    \mathcal O_{q,x}^{i}{}_{\beta}
    G_{R,q}^{\beta}{}_{j}
    =
    i\delta'(\tau)
    \mathcal C_q^{i}{}_{j}
    +
    \delta(\tau)
    \left[
        2i\partial_t
        \mathcal C_q^{i}{}_{j}
        -
        \Pi_q^i
        \mathcal C_q^{0}{}_{j}
    \right].
\end{equation}

Using the distributional identity
\begin{align}
    \mathcal C_q^{i}{}_{j}(t,t')
    \delta'(\tau)
    =
    \left.
    \mathcal C_q^{i}{}_{j}
    \right|_{t=t'}
    \delta'(\tau)
    -
    \left.
    \partial_t
    \mathcal C_q^{i}{}_{j}
    \right|_{t=t'}
    \delta(\tau),
\end{align}
which, since the source term of the Green equation contains no $\delta'(\tau)$ term, leads to the vanishing of the purely spatial equal-time Pauli-Jordan function
\begin{equation}
    \left.
    \mathcal C_q^{i}{}_{j} (x,x')
    \right|_{t=t'}
    =
    0.
\end{equation}
such that the identity simplifies to
\begin{equation}
    i\delta'(\tau)\,
    \mathcal C_q^{i}{}_{j}
    =
    -
    i\delta(\tau)
    \left.
    \partial_t
    \mathcal C_q^{i}{}_{j}
    \right|_{t=t'}.
\end{equation}
When plugged back into the Green equation, 
\begin{equation}
    \mathcal O_{q,x}^{i}{}_{\beta}\,
    G_{R,q}^{\beta}{}_{j}
    =
    \delta(\tau)
    \left.
    \left[
        i\partial_t
        \mathcal C_q^{i}{}_{j}
        -
        \Pi_q^i
        \mathcal C_q^{0}{}_{j}
    \right]
    \right|_{t=t'},
\end{equation}
evaluated at $t = t'$, enforced by the delta distribution.

Putting everything together, this reads
\begin{equation}
\begin{aligned}
    \left.
    \left[
        i\partial_t
        \mathcal C_q^{i}{}_{j}(x,x')
        -
        \Pi_q^i
        \mathcal C_q^{0}{}_{j}(x,x')
    \right]
    \right|_{t=t'}\\
    =
    -
    \delta^i{}_{j}
    \delta^{(3)}(\boldsymbol x-\boldsymbol x').
\end{aligned}
\label{eq: completeness_condition_precursor}
\end{equation}

Since $\partial_t$ and $\Pi_q^i$ act on the first argument $x$, one
obtains
\begin{equation}
\begin{aligned}
    &i\partial_t
    \mathcal C_q^{i}{}_{j}(x,x')
    -
    \Pi_q^i
    \mathcal C_q^{0}{}_{j}(x,x')
    \\
    &\qquad= - i
    \sum_{\lambda=1}^{3}
    \int d\Gamma_p
    \sum_{r=\pm1}r\,
    \varpi_{q,r,\lambda}^{i}(\boldsymbol p, x)
    \bar w_{q,r,\lambda,j}(\boldsymbol p, x'),
\end{aligned}
\end{equation}
where we introduced the canonical momentum
\begin{equation}
    \label{eq: canonical_momentum_mode}
    - i \varpi_{q,r,\lambda}^{i}(\boldsymbol p,x)
    \coloneqq
    i\partial_t
    w_{q,r,\lambda}^{i}(\boldsymbol p,x)
    -
    \Pi_q^i
    w_{q,r,\lambda}^{0}(\boldsymbol p,x),
\end{equation}
which, expressed in terms of the field strength tensor, reads
\begin{equation}
    \varpi_{q,r,\lambda}^{i}
    =
    - \left(
    D_q^0 w_{q,r,\lambda}^{i}
    -
    D_q^i w_{q,r,\lambda}^{0}
    \right)
    \eqqcolon
    -
    \mathcal W_{q,r,\lambda}^{0i},
\end{equation}
using the gauge $\mathcal{A}^0 = 0$, where $\Pi^0_q = i \partial_t$ and $D_q^\mu=-i\Pi_q^\mu$.

Hence, the Green equation involving the retarded propagator lead to the completeness relation for the modes
\begin{equation}
\begin{aligned}
    i\sum_{\lambda=1}^{3}\int d\Gamma_p
    \sum_{r=\pm1} r\,
    \left.
    \varpi_{q,r,\lambda}^{i}(\boldsymbol p,x)\,
    \bar w_{q,r,\lambda,j}(\boldsymbol p,x')
    \right|_{t=t'}
    \\
    =
    \delta^i{}_{j}\,
    \delta^{(3)}(\boldsymbol x-\boldsymbol x'),
\end{aligned}
\label{eq:completeness_relation_VP_modes}
\end{equation}
and to the vanishing of the equal-time Pauli-Jordan function
\begin{equation}
    \sum_{\lambda=1}^{3}
    \int d\Gamma_p
    \sum_{r=\pm1}r\,
    \left.
    w_{q,r,\lambda}^{i}(\boldsymbol p,x)\,
    \bar w_{q,r,\lambda,j}(\boldsymbol p,x')
    \right|_{t=t'}
    = 0,
\end{equation}
in close analogy with Ref.~\cite{DiPiazza:2018ofz}, where the completeness relation for Volkov fermions is derived from the Feynman propagator.

\section{Quantization of the theory}
\label{sec: quantization}

\subsection{Canonical quantization for fields}

The canonical quantization of the theory proceeds by promoting the mode
coefficients in Eq.~\eqref{eq: Volkov_Proca_on_shell_expansion} to
operators,
\begin{equation}
    c_q^r(\boldsymbol p)
    \;\longmapsto\;
    \hat c_q^r(\boldsymbol p),
    \qquad r=\pm,
\end{equation}
such that the quantized Volkov--Proca field is given by
\begin{equation}
    \label{eq: Volkov_Proca_quantized_expansion}
    \hat W_q^\mu(x)
    =
    \int
    d\Gamma_p\,
    \sum_{r=\pm}
    \hat c_q^r(\boldsymbol p)\,
    w_q^{r\,\mu}(\boldsymbol p,x),
\end{equation}
under the identifications
\begin{equation}\label{eq: CreaAnnh}
    \hat c_q^+(\boldsymbol p)
    \equiv
    \hat a_q(\boldsymbol p),
    \qquad
    \hat c_q^-(\boldsymbol p)
    \equiv
    \hat b_{-q}^{\dagger}(\boldsymbol p),
\end{equation}
with corresponding annihilation and creation operators.

The corresponding canonical-momentum modes are
\begin{equation}
    \varpi_{q}^{r\,i}(\boldsymbol p,x)
    \coloneqq
    -\mathcal W_{q}^{r\,0i}(\boldsymbol p,x),
    \qquad r=\pm,
\end{equation}
such that the canonical momentum field\footnote{Should not be confused with the dressed momentum operator.} admits the expansion
\begin{equation}
    \hat\Pi_q^i(x)
    =
    \int
    \frac{d^3p}{(2\pi)^3\,2\varepsilon_{\boldsymbol p}}\,
    \sum_{r=\pm}
    \hat c_{q}^r(\boldsymbol p)\,
    \varpi_{q}^{r\,i}(\boldsymbol p,x).
    \label{eq:Volkov_Proca_momentum_expansion}
\end{equation}
again under $\hat c_{q}^{+}(\boldsymbol p) \equiv \hat a_{q}(\boldsymbol p)$ and $\hat c_{q}^{-}(\boldsymbol p) \equiv \hat b_{-q}^{\dagger}(\boldsymbol p)$.

Since the temporal canonical momentum vanishes identically,
$\hat\Pi_q^0=0$, the independent canonical variables are the spatial
components. The equal-time canonical commutation relations between the canonically conjugated variables are then
\begin{align}
    \left[
        \hat W_q^i(t,\boldsymbol x),
        \hat\Pi^\dagger_{q,j}(t,\boldsymbol y)
    \right]
    &=
    -i\delta^i{}_{j}\,
    \delta^{(3)}(\boldsymbol x-\boldsymbol y),
    \label{eq:VP_canonical_field_commutator}
\end{align}
where the dagger on the canonical momentum is due to the fact that under this notation, $q$ represents the charge of the resulting object, but the canonical conjugate to $\hat{W}_q^i(x)$ is actually $\hat\Pi^\dagger_{q,j} = \hat\Pi_{-q,j}$, which can be seen in Eq.~\eqref{eq: canonical_energy_momentum_tensor} and from
\begin{equation}
\begin{aligned}
    \mathsf{\Pi}_{- q}^{\rho\alpha} (x)
    &\coloneqq
    \frac{\partial\mathcal L_{\mathrm{VP}}}
    {\partial\!\left(
        \partial_\rho
        W_{q,\alpha} (x)
    \right)} =
    -\mathcal W_{-q}^{\rho\alpha} (x),
\end{aligned}
\end{equation}
by taking the $0-$component, i.e. $\Pi^\mu_q (x) = \mathsf{\Pi}_{q}^{0\mu} (x)$.

\subsection{Canonical quantization in the polarization basis}

To make the three physical polarization degrees of freedom explicit, we
adopt a basis $\{\lambda\}_{\lambda=1}^{3}$ of polarization states and
resolve the mode functions and their coefficients accordingly,
\begin{equation}
    w_q^{r\,\mu}(\boldsymbol p,x)
    \;\mapsto\;
    w_{q,\lambda}^{r\,\mu}(\boldsymbol p,x),
    \quad
    c_q^r(\boldsymbol p)
    \;\mapsto\;
    c_{q,\lambda}^r(\boldsymbol p),
\end{equation}
such that the field expansion takes the form
\begin{equation} \label{eq: quantized_field}
    \hat W_q^\mu(x)
    =
    \sum_{\lambda=1}^{3}
    \int
    d\Gamma_p\,
    \sum_{r=\pm}
    \hat c_{q,\lambda}^r(\boldsymbol p)\,
    w_{q,\lambda}^{r\,\mu}(\boldsymbol p,x),
\end{equation}
with $\hat c_{q,\lambda}^{+}(\boldsymbol p) \equiv \hat a_{q,\lambda}(\boldsymbol p)$ and $\hat c_{q,\lambda}^{-}(\boldsymbol p) \equiv \hat b_{-q,\lambda}^\dagger (\boldsymbol p)$.

The mode coefficients for the quantized fields still satisfy Eq.~\eqref{eq: mode_coefficients_innner_product}, but now as operators, which reads in split form
\begin{equation} 
\label{eq: creation_annihilation_ops}
    \begin{aligned}
        \hat a_{q,\lambda}(\boldsymbol p)
        &=
        \bigl(
            w_{q,+,\lambda}(\boldsymbol p),
            \hat W_q
        \bigr),
        \\
        \hat b_{-q,\lambda}^{\dagger}(\boldsymbol p)
        &=
        -
        \bigl(
            w_{q,-,\lambda}(\boldsymbol p),
            \hat W_q
        \bigr).
    \end{aligned}
\end{equation}

The nonvanishing commutation relations of the creation and annihilation operators are chosen as
\small
\begin{align}
    \left[
        \hat a_{q,\lambda}(\boldsymbol p),
        \hat a_{q,\lambda'}^\dagger(\boldsymbol p')
    \right]
    &=
    (2\pi)^3\,2\varepsilon_{\boldsymbol p}\,
    \delta_{\lambda\lambda'}\,
    \delta^{(3)}(\boldsymbol p-\boldsymbol p'),
    \\
    \left[
        \hat b_{-q,\lambda}(\boldsymbol p),
        \hat b_{-q,\lambda'}^\dagger(\boldsymbol p')
    \right]
    &=
    (2\pi)^3\,2\varepsilon_{\boldsymbol p}\,
    \delta_{\lambda\lambda'}\,
    \delta^{(3)}(\boldsymbol p-\boldsymbol p'),
\end{align}
\normalsize
while all crossed commutators vanish,
\begin{equation}
    \left[
        \hat a_{q,\lambda},
        \hat b_{-q,\lambda'}
    \right]
    =
    \left[
        \hat a_{q,\lambda},
        \hat b_{-q,\lambda'}^\dagger
    \right]
    =
    0,
\end{equation}
as well as commutators among two same type operators.

\subsection{Quantum field representation for the propagators}

We now express the propagating parts of the retarded, advanced, and
Feynman Green functions in terms of the quantized Volkov--Proca field.
The following representations are understood away from coincident
spacetime points, $x\neq x'$. Indeed, the full Green functions contain,
in addition to their on-shell propagating contributions, the local
contact term derived in Eq.~\eqref{eq: dressed_Proca_contact_term},
whose support is restricted to $x=x'$. Consequently, for $x\neq x'$
the full Green functions coincide with their propagating parts, and we
omit the corresponding ``$\mathrm{dec}$'' subscript in what follows.

It is useful to introduce the Wightman functions associated with the
positive- and negative-frequency sectors,
\begin{equation}
    \mathcal{G}_{q,r}^{\mu\nu}(x,x')
    =
    \sum_{\lambda=1}^{3}
    \int d\Gamma_p\,
    w_{q,\lambda}^{r\,\mu}(\boldsymbol p,x)\,
    \bar w_{q,\lambda}^{r\,\nu}(\boldsymbol p,x'),
    \label{eq: Wightman_functions_Volkov_Proca}
\end{equation}
which correspond to the positive- and negative-frequency propagating sectors,
respectively. In terms of the quantized fields, with
$\hat W_{-q}=\hat W_q^\dagger$, they are written as
\begin{align}
\mathcal{G}_{q,+}^{\mu\nu}(x,x')
&=
\bra{\Omega}
\hat W_q^\mu(x)\,
\hat W_{-q}^\nu(x')
\ket{\Omega},
\\
\mathcal{G}_{q,-}^{\mu\nu}(x,x')
&=
\bra{\Omega}
\hat W_{-q}^\nu(x')\,
\hat W_q^\mu(x)
\ket{\Omega},
\label{eq: Wightman_functions_Volkov_Proca_field}
\end{align}
such that the Feynman propagator reads
\begin{equation}
    \begin{aligned}
        iG_{F,q}^{\mu\nu}(x,x')
        &=
        \sum_{r=\pm}
        \theta(r\tau)\,
        \mathcal{G}_{q,r}^{\mu\nu}(x,x'),
        \qquad x \neq x'
    \end{aligned}
    \label{eq: Feynman_propagator_Wightman}
\end{equation}
where $\tau=t-t'$ and similarly,
\begin{align}
    \mathcal{C}_q^{\mu\nu}(x,x')
    &=
    \sum_{r=\pm} r\,
    \mathcal{G}_{q,r}^{\mu\nu}(x,x')
    \notag\\
    &=
    \bra{\Omega}
    \left[
        \hat W_q^\mu(x),
        \hat W_{-q}^{\nu}(x')
    \right]
    \ket{\Omega},
    \label{eq: Pauli_Jordan_Wightman}
\end{align}
for the Pauli--Jordan function\footnote{The expression of the Pauli--Jordan function in terms of the commutator justifies the notation by $\mathcal{C}^{\mu\nu}$.}, such that
\begin{equation}
\begin{aligned}
    iG_{R,q}^{\mu\nu}(x,x')
    &=
    \theta(\tau)\,
    \mathcal C_q^{\mu\nu}(x,x'),
    \\
    iG_{A,q}^{\mu\nu}(x,x')
    &=
    -\theta(-\tau)\,
    \mathcal C_q^{\mu\nu}(x,x'),
\end{aligned}
\qquad x\neq x'.
\label{eq: retarded_advanced_Jordan}
\end{equation}
for the advanced and retarded propagators.

Expanding out the sum in Eq.~\eqref{eq: Feynman_propagator_Wightman} leads to the familiar expression \cite{Weinberg:1995}, the starting point in works like \cite{Monreal5jyf-llwf},
\begin{equation}
    iG_{F,q}^{\mu\nu}(x,x')
    =
    \bra{\Omega}
    \mathcal T
    \left\{
        \hat W_q^\mu(x)\,
        \hat W_{-q}^\nu(x')
    \right\}
    \ket{\Omega},
    \qquad x\neq x',
\label{eq: Feynman_propagator_time_ordered}
\end{equation}
where $\mathcal T$ denotes the time-ordering operator, such that
\begin{equation}
\begin{aligned}
    \mathcal T
    \left\{
        \hat W_q^\mu(x)\,
        \hat W_{-q}^\nu(x')
    \right\}
    &=
    \theta(\tau)\,
    \hat W_q^\mu(x)\,
    \hat W_{-q}^\nu(x')
    \\
    &+
    \theta(-\tau)\,
    \hat W_{-q}^\nu(x')\,
    \hat W_q^\mu(x),
\end{aligned}
\label{eq: time_ordering_Volkov_Proca}
\end{equation}
where again $\tau=t-t'$ is the time difference.

\subsection{Frequency sectors projectors}

The projectors onto the positive- and negative-frequency subspaces of the
solution space are defined by
\begin{equation}
    \mathsf P_{q,r} W_q
    \coloneqq
    r
    \sum_{\lambda=1}^{3}
    \int d\Gamma_p\,
    w_{q,\lambda}^{r}(\boldsymbol p)\,
    \bigl(
        w_{q,\lambda}^{r}(\boldsymbol p),
        W_q
    \bigr),
    \label{eq: frequency_projectors}
\end{equation}
where the factor $r$ compensates for the indefinite normalization of the
positive- and negative-frequency modes, as in Eq.~\eqref{eq: orthogonality_relations_classical}. Consequently, they pick out the corresponding mode
\begin{equation}
    \mathsf P_{q,r}\,
    w_{q,\lambda}^{r'}(\boldsymbol p)
    =
    \delta_{rr'}\,
    w_{q,\lambda}^{r'}(\boldsymbol p),
\end{equation}
and satisfy the projector identities
\begin{equation}
    \mathsf P_{q,r}\mathsf P_{q,r'}
    =
    \delta_{rr'}\,\mathsf P_{q,r},
    \qquad
    \sum_{r=\pm}\mathsf P_{q,r}
    =
    \mathbb 1.
\end{equation}

Although introduced here in the context of canonical quantization, the
operators $\mathsf P_{q,r}$ act on the classical solution space, rather than directly on the Fock space, and hats are therefore omitted from their notation. Upon quantization, the frequency-sector projectors induce the corresponding projectors on the $1$-particle Hilbert space, while on Fock space they give rise to the associated particle- and antiparticle-number operators. The projector identities above therefore refer to the $1$-particle sector.

\subsection{Conserved charges}
\label{subsec: conserved_charges}

\subsubsection{Global \texorpdfstring{$U(1)$}{U(1)} charge}

The conserved electric charge in Eq.~\eqref{eq: conserved_charge_classical} is quantized to 
\begin{equation}
\begin{aligned}
    :
    \widehat{\mathcal Q}_{\mathrm{el}}
    :
    &=
    q e\,
    :
    \bigl(\widehat W_q,\widehat W_q\bigr)
    :\\
    &=
    q e
    \sum_{\lambda=1}^{3}
    \int d\Gamma_p\,
    \sum_{r=\pm1}
    r\,
    \widehat N_{q,r,\lambda}(\boldsymbol p),
\end{aligned}
\end{equation}
where the momentum-space number operators are
\begin{equation}
    \widehat N_{q,r,\lambda}(\boldsymbol p)
    \coloneqq
    \begin{cases}
        \hat a_{q,\lambda}^{\dagger}(\boldsymbol p)\,
        \hat a_{q,\lambda}(\boldsymbol p),
        & r=+,
        \\[1mm]
        \hat b_{-q,\lambda}^{\dagger}(\boldsymbol p)\,
        \hat b_{-q,\lambda}(\boldsymbol p),
        & r=-.
    \end{cases}
\end{equation}
for particles and anti-particles, respectively.

\subsubsection{Residual translation charges}

For a translation generated by a vector $\xi^\mu$, satisfying
$\xi\cdot n = 0$, the corresponding conserved Noether charge is
\begin{equation}
    :
    \xi\cdot\widehat P_{\mathrm{can}}
    :
    =
    \sum_{\lambda=1}^{3}
    \int d\Gamma_p\,
    (\xi\cdot p)\,
    \widehat N_{q,\lambda}(\boldsymbol p),
    \label{eq:conserved_spacetime_charges}
\end{equation}
where the total momentum-space number operator is
\begin{equation}
    \widehat N_{q,\lambda}(\boldsymbol p)
    \coloneqq
    \widehat N_{q,+,\lambda}(\boldsymbol p)
    +
    \widehat N_{q,-,\lambda}(\boldsymbol p).
\end{equation}

Spacetime generators not belonging to the residual translation group in Eq.~\eqref{eq: residual_connected_translation_group} are not diagonal in the Volkov--Proca basis and generally contain terms mixing creation and annihilation operators.

\section{LSZ reduction formula for charged vector bosons in a plane-wave field}
\label{sec: LSZ_reduction}

As an application of the Volkov--Proca states constructed above, we
derive the Lehmann--Symanzik--Zimmermann (LSZ) reduction formula \cite{Lehmann:1954rq} for
charged vector bosons propagating in an electromagnetic plane-wave
background. The external asymptotic states are described by Volkov--Proca modes rather than by free plane-wave modes, so that the interaction with the external background is incorporated exactly, while the remaining dynamical interactions are treated through the interacting Heisenberg field\footnote{To be distinguished from the free Volkov--Proca field, denoted the same.} $\hat W_q^\mu(x)$, which satisfies the sourced equation of motion
\begin{equation}
    \left(
        \mathcal O_q
    \right)^\mu{}_\nu
    \hat W_q^\nu(x)
    =
    -\hat J_q^\mu(x),
    \label{eq: Heisenberg_field_EOM}
\end{equation}
where $\mathcal O_q$ is the background-dressed Volkov--Proca operator defined in Eq.~\eqref{def: VP_EOM_operator}, and $\hat J_q^\mu$ collects the dynamical interaction terms not included in the electromagnetic background.

\subsection{Asymptotic fields}

We assume the existence of asymptotic field operators,
\[
    \hat W_q^{\mu,\mathrm{in/out}}(x),
\]
associated with the interacting field $\hat W_q^\mu(x)$, such that
\begin{equation} 
    \underset{t\to\mp\infty}{\mathrm{w}\!-\!\lim}\, Z_W^{-1/2}\, \hat W_q^\mu(x) = \hat W_q^{\mu,\mathrm{in/out}}(x), 
    \label{eq: LSZ_asymptotic_field_condition} 
\end{equation}
where $Z_W$ denotes the wave-function normalization factor and with the limit understood in the weak sense, namely at the level of matrix elements between suitable states,\footnote{More precisely, since quantum fields are operator-valued distributions, the asymptotic condition is understood after smearing with appropriate test functions.} as in the standard LSZ construction \cite{Haag:1996}.

Provided that the action of the conserved generators associated with the symmetries preserved by the prescribed background can be interchanged with the weak asymptotic limit in Eq.~\eqref{eq: LSZ_asymptotic_field_condition}, the asymptotic fields inherit the corresponding transformation properties of the interacting field. In particular, for the continuous residual translations generated by $\widehat{\mathcal Q}_\xi$, with $\xi\cdot n=0$, the asymptotic fields transform according to
\begin{equation}
    \left[
        \widehat{\mathcal Q}_\xi,
        \hat W_q^{\mu,\mathrm{in/out}}(x)
    \right]
    =
    -
    i\,\xi\cdot\partial\,
    \hat W_q^{\mu,\mathrm{in/out}}(x),
    \label{eq: LSZ_residual_translation_asymptotic}
\end{equation}

The asymptotic fields are free with respect to the remaining
dynamical interactions, while their coupling to the prescribed
electromagnetic background is incorporated exactly. Accordingly,
they satisfy the free Volkov--Proca equation
\begin{equation}
    \left(
        \mathcal O_q
    \right)^\mu{}_\nu\,
    \hat W_q^{\nu,\mathrm{in/out}}(x)
    =
    0,
\end{equation}
with the physical $W$-boson mass entering the equation.

\subsection{Asymptotic mode operators}

Integrating the current divergence identity from Eq.~\eqref{eq: current_divergence_EOM},
\begin{equation}
    \partial_\mu j^\mu(U,V)
    =
    i\,
    \bar U_\mu
    (
        \overleftrightarrow{\mathcal O}_{Q}
    )^\mu{}_\nu
    V^\nu,
    \label{eq: LSZ_local_Green_identity}
\end{equation}
for two fields belonging to the same charge sector $q$, over the spacetime slab bounded by the hypersurfaces $\Sigma_{t_1}$ and $\Sigma_{t_2}$,
\begin{equation}
    (U,V)_{t_2}
    -
    (U,V)_{t_1}
    =
    i
    \int_{\Omega_{12}} d^4x\,
    \bar U_\mu
    (
        \overleftrightarrow{\mathcal O}_{Q}
    )^\mu{}_\nu
    V^\nu,
    \label{eq: LSZ_integrated_Green_identity}
\end{equation}
since the spatial boundary contribution vanishing for localized fields. Using that for a free Volkov--Proca mode,
\begin{equation}
    (\mathcal O_q)^\mu{}_\nu\,
    w_{q,r,\lambda}^{\nu}(\boldsymbol p,x)
    =
    0,
\label{eq: LSZ_mode_EOM}
\end{equation}
and its complex conjugate satisfies
\begin{equation}
    \bar w_{q,r,\lambda,\mu}(\boldsymbol p,x)
    (\overleftarrow{\mathcal O}_{-q})^\mu{}_\nu
    =
    0,
\label{eq: LSZ_conjugate_mode_EOM}
\end{equation}
by taking $t_1\to-\infty$ and $t_2\to+\infty$, and setting
$U=w_{q,r,\lambda}$ and $V=\hat W_q$ in
Eq.~\eqref{eq: LSZ_integrated_Green_identity}, one obtains
\begin{equation}
\begin{aligned}
    &
    \left(
        w_{q,r,\lambda},
        \hat W_q
    \right)_{\mathrm{out}}
    -
    \left(
        w_{q,r,\lambda},
        \hat W_q
    \right)_{\mathrm{in}}
    \\
    &\qquad
    =
    i\int d^4x\,
    \bar w_{q,r,\lambda,\mu}(\boldsymbol p,x)
    (\mathcal O_q)^\mu{}_\nu
    \hat W_q^\nu(x).
\end{aligned}
\label{eq: LSZ_mode_field_difference}
\end{equation}

The projection of the interacting field onto the free modes\footnote{Equation~\eqref{eq: LSZ_mode_extraction} defines the
time-dependent projection of the full interacting field onto the
free Volkov--Proca modes. No expansion of the interacting field
in this mode basis is assumed at finite times. Such an expansion would
require the Volkov--Proca modes to form a complete basis for the
interacting Cauchy data, including the corresponding constraints.
Only in the asymptotic regions, where the interacting field approaches
the asymptotic Volkov--Proca field, do these projections become the
creation and annihilation operators appearing in its mode expansion.},
\begin{equation}
    \hat c_{q,\lambda}^{r}(\boldsymbol p, t)
    \coloneqq
    Z_W^{- 1 / 2}
    r\left(
        w_{q,r,\lambda}(\boldsymbol p),
        \hat W_q
    \right)_t,
    \label{eq: LSZ_mode_extraction}
\end{equation}
allows to define the mode operators at asymptotic times by
\begin{equation}
\begin{aligned}
    \hat c_{q,\lambda}^{r,\mathrm{out}}(\boldsymbol p)
    &\coloneqq
    \hat c_{q,\lambda}^{r}(\boldsymbol p,t=+\infty),
    \\
    \hat c_{q,\lambda}^{r,\mathrm{in}}(\boldsymbol p)
    &\coloneqq
    \hat c_{q,\lambda}^{r}(\boldsymbol p,t=-\infty),
\end{aligned}
\qquad
r=\pm,
\label{eq: LSZ_asymptotic_mode_operators}
\end{equation}
with $\hat c_{q,\lambda}^{+} = \hat a_{q,\lambda}$ and $\hat c_{q,\lambda}^{-} = \hat b_{-q,\lambda}^{\dagger}$, which, under
\begin{equation}
    \hat a_{q,\lambda}^{\mathrm{in/out}}(\boldsymbol p)
    \ket{\Omega}
    =
    \hat b_{-q,\lambda}^{\mathrm{in/out}}(\boldsymbol p)
    \ket{\Omega}
    =
    0,
\end{equation}
defines the interacting vacuum $\ket{\Omega}$. Thus, Eq.~\eqref{eq: LSZ_mode_field_difference} yields
\begin{equation}
\begin{aligned}
    &\hat c_{q,\lambda}^{r,\mathrm{out}}(\boldsymbol p)
    -
    \hat c_{q,\lambda}^{r,\mathrm{in}}(\boldsymbol p)
    \\
    &\qquad
    =
    Z_W^{- 1 / 2}
    ir
    \int d^4x\,
    \bar w_{q,r,\lambda,\mu}(\boldsymbol p,x)
    \left(
        \overrightarrow{\mathcal O}_{q}
    \right)^\mu{}_\nu
    \hat W_q^\nu(x).
\end{aligned}
\label{eq: LSZ_c_difference}
\end{equation}

Notice that taking the weak limit from Eq.~\eqref{eq: LSZ_asymptotic_field_condition} leads to
\begin{equation}
    \hat c_{q,\lambda}^{r,\mathrm{in/out}}(\boldsymbol p)
    =
    r\left(
        w_{q,r,\lambda}(\boldsymbol p),
        \hat W_q^{\mathrm{in/out}}
    \right)_\mathrm{in/out},
\end{equation}
so that the asymptotic limit of the projection from Eq.~\eqref{eq: LSZ_asymptotic_mode_operators} of the interacting field provides actually the creation and annihilation operators appearing in the mode expansion of the (free) asymptotic Volkov--Proca field.

The adjoint relations follow from the hermiticity property
\begin{equation}
    (U,V)^\dagger=(V,U),
\end{equation}
which, together with Eq.~\eqref{eq: LSZ_mode_extraction}, leads to the projection
\begin{equation}
    \hat c_{q,\lambda}^{r\,\dagger}(\boldsymbol p, t)
    =
    Z_W^{- 1 / 2}
    r\left(
        \hat W_q,
        w_{q,r,\lambda}(\boldsymbol p)
    \right)_t,
    \label{eq: LSZ_adjoint_mode_extraction}
\end{equation}
where $\hat c_{q,\lambda}^{+\,\dagger} = \hat a_{q,\lambda}^{\dagger}$ and $\hat c_{q,\lambda}^{-\,\dagger} = \hat b_{-q,\lambda}$. Taking instead $U=\hat W_q$ and
$V=w_{q,r,\lambda}$ in Eq.~\eqref{eq: LSZ_integrated_Green_identity}, the first term on the
right-hand side vanishes by Eq.~\eqref{eq: LSZ_mode_EOM}, yielding
\begin{equation}
\begin{aligned}
    &
    \left(
        \hat W_q,
        w_{q,r,\lambda}
    \right)_{\mathrm{out}}
    -
    \left(
        \hat W_q,
        w_{q,r,\lambda}
    \right)_{\mathrm{in}}
    \\
    &\qquad
    =
    - i\int d^4x\,
    \hat W_{-q,\mu}(x)
    (\overleftarrow{\mathcal O}_{-q})^\mu{}_\nu
    w_{q,r,\lambda}^{\nu}(\boldsymbol p,x),
\end{aligned}
\label{eq: LSZ_adjoint_mode_field_difference}
\end{equation}
where $\hat W_{-q}=\hat W_q^\dagger$. Therefore, using
Eq.~\eqref{eq: LSZ_adjoint_mode_extraction}, the two frequency sectors
can be written in the common form
\begin{equation}
\begin{aligned}
    &\hat c_{q,\lambda}^{r\,\dagger,\mathrm{out}}(\boldsymbol p)
    -
    \hat c_{q,\lambda}^{r\,\dagger,\mathrm{in}}(\boldsymbol p)
    \\
    &\qquad
    =
    - Z_W^{- 1 / 2} i r
    \\
    &\qquad\quad\times
    \int d^4x\,
    \hat W_{-q,\mu}(x)
    \left(
        \overleftarrow{\mathcal O}_{-q}
    \right)^\mu{}_\nu
    w_{q,r,\lambda}^{\nu}(\boldsymbol p,x).
\end{aligned}
\label{eq: LSZ_cdagger_difference}
\end{equation}

\subsection{LSZ reduction of the transition amplitude}

We now consider asymptotic multiparticle states built on the full vacuum $\ket{\Omega}$ of the interacting Volkov--Proca theory in the presence of the prescribed plane-wave background. The initial state, containing $n$ charged bosons of charge $q$ and $m$ bosons of charge $-q$, is
\begin{equation}
\begin{aligned}
    \ket{i}
    =
    \left[
        \prod_{k=1}^{n}
        \hat a_{q,\lambda_k}^{\dagger\,\mathrm{in}}
        (\boldsymbol p_k)
    \right]
    \left[
        \prod_{\ell=1}^{m}
        \hat b_{-q,\rho_\ell}^{\dagger\,\mathrm{in}}
        (\boldsymbol q_\ell)
    \right]
    \ket{\Omega},
\end{aligned}
\label{eq: LSZ_initial_state}
\end{equation}
and a final state containing $n'$ bosons of charge $q$ and $m'$ bosons
of charge $-q$,
\begin{equation}
\begin{aligned}
    \bra{f}
    =
    \bra{\Omega}
    \left[
        \prod_{i=1}^{n'}
        \hat a_{q,\lambda_i'}^{\mathrm{out}}
        (\boldsymbol p_i')
    \right]
    \left[
        \prod_{j=1}^{m'}
        \hat b_{-q,\rho_j'}^{\mathrm{out}}
        (\boldsymbol q_j')
    \right].
\end{aligned}
\label{eq: LSZ_final_state}
\end{equation}

The reduction is performed recursively. For an outgoing particle one
retains the difference
$\hat a^{\mathrm{out}}-\hat a^{\mathrm{in}}$, whereas for an incoming
particle the connected contribution contains
$\hat a^{\dagger\,\mathrm{in}}-\hat a^{\dagger\,\mathrm{out}}$.
The terms obtained by commuting an in-operator through the initial state,
or an out-operator through the final state, generate the usual
disconnected contributions. After separating these terms, each external
leg of the connected transition amplitude contributes the same overall
factor $i$. Defining the total number of particles and anti-particles as
\begin{equation}
    N=n+m+n'+m',
\end{equation}
the resulting background-dressed LSZ reduction formula is
\begin{widetext}
\begin{equation}
\begin{aligned}
    \bra{f}S\ket{i}_{\mathrm{conn}}
    ={}
    i^{N} Z_W^{-N/2}
    &\times 
    \prod_{i=1}^{n'}
    \int d^4x_i\,
    \bar w_{q,+,\lambda_i',\mu_i}
    (\boldsymbol p_i',x_i)
    \left(
        \overrightarrow{\mathcal O}_{q,x_i}
    \right)^{\mu_i}{}_{\alpha_i}
    \\
    &\times
    \prod_{\ell=1}^{m}
    \int d^4u_\ell\,
    \bar w_{q,-,\rho_\ell,\nu_\ell}
    (\boldsymbol q_\ell,u_\ell)
    \left(
        \overrightarrow{\mathcal O}_{q,u_\ell}
    \right)^{\nu_\ell}{}_{\beta_\ell}
    \\
    &\times
    \bra{\Omega}
    \mathcal T
    \Bigg\{
        \prod_{i=1}^{n'}
        \hat W_q^{\alpha_i}(x_i)
        \prod_{j=1}^{m'}
        \hat W_{-q}^{\sigma_j}(y_j)
        \prod_{k=1}^{n}
        \hat W_{-q}^{\gamma_k}(z_k)
        \prod_{\ell=1}^{m}
        \hat W_q^{\beta_\ell}(u_\ell)
    \Bigg\}
    \ket{\Omega}_{\mathrm{conn}}
    \\
    &\times
    \prod_{j=1}^{m'}
    \int d^4y_j\,
    \left(
        \overleftarrow{\mathcal O}_{-q,y_j}
    \right)_{\sigma_j\tau_j}
    w_{q,-,\rho_j'}^{\tau_j}
    (\boldsymbol q_j',y_j)
    \\
    &\times
    \prod_{k=1}^{n}
    \int d^4z_k\,
    \left(
        \overleftarrow{\mathcal O}_{-q,z_k}
    \right)_{\gamma_k\delta_k}
    w_{q,+,\lambda_k}^{\delta_k}
    (\boldsymbol p_k,z_k).
\end{aligned}
\label{eq: LSZ}
\end{equation}
\end{widetext}
Here, the arrows specify the direction in which the corresponding EOM
operators act: $\overrightarrow{\mathcal O}$ acts on the time-ordered
Green function to its right, whereas $\overleftarrow{\mathcal O}$ acts
on it to its left. Equation~\eqref{eq: LSZ}
thus amputates each external propagator with the inverse Volkov--Proca
operator and replaces the corresponding free external wave function by
the appropriate Volkov--Proca mode.

Equation~\eqref{eq: LSZ} is the plane-wave analogue of the LSZ
construction for charged vector bosons in a constant magnetic
background \cite{Monreal5jyf-llwf}. The effect of the external
plane-wave field is encoded both in the Volkov--Proca external states
and in the background-dependent EOM operators. In the vacuum limit,
the Volkov--Proca modes reduce to the ordinary Proca plane waves and
$\mathcal O_q$ reduces to the free Proca inverse propagator, recovering
the standard LSZ reduction formula for a massive complex vector field.

\section{K\"all\'en--Lehmann spectral representation for the Volkov--Proca field}
\label{sec: Kallen-Lehmann_spectral}

The ordinary K\"all\'en--Lehmann representation \cite{Kallen:1952zz, Lehmann:1954xi} cannot be derived unchanged in a Volkov background, because the background breaks the full translation group. Instead, what can be derived is a spectral representation with respect to the residual translation group from Eq.~\eqref{eq: residual_connected_translation_group}. Since there are only three independent continuous translation generators rather than four, exact states cannot in general be classified by four conserved momentum components. Instead, we shall use 
\[
\ket{\alpha,\boldsymbol\kappa},
\]
a complete set of exact Hilbert-space states of the full interacting theory in the external Volkov background, with 
\begin{equation}
\boldsymbol{\kappa} = \left(n \cdot p, a_1 \cdot p, a_2 \cdot p\right) 
\equiv
    \left(
        p^+,
        p^{\perp,1},
        p^{\perp,2}
    \right),
\end{equation}
the momentum $p$ projected along $n^\perp \equiv \operatorname{span} \{ n, a_1, a_2 \}$, to denote a complete set of simultaneous generalized eigenstates of the three commuting residual translation generators $Q_A$, $A=1,2,3$, satisfying
\begin{equation}
    Q_A\ket{\alpha,\boldsymbol\kappa}
    =
    \kappa_A\ket{\alpha,\boldsymbol\kappa},
    \label{eq: residual_momentum_eigenstates}
\end{equation}
with $\alpha$ collecting the remaining quantum numbers. For any residual translation $\xi\in T_{\mathcal A}^{(0)}$, the corresponding unitary operator acts on these states according to
\begin{equation}
    U(\xi)\ket{\alpha,\boldsymbol\kappa}
    =
    e^{i\kappa(\xi)}
    \ket{\alpha,\boldsymbol\kappa},
    \qquad
    \kappa(\xi)
    \coloneqq
    p\cdot\xi,
    \label{eq: residual_translation_state_action}
\end{equation}
thus, under this action, the covariance of the field reads
\begin{equation}
\begin{aligned}
    \bra{\Omega}
    \hat W_q^\mu(x+\xi)
    \ket{\alpha,\boldsymbol\kappa}
    &=
    e^{-i\kappa(\xi)}
    \bra{\Omega}
    \hat W_q^\mu(x)
    \ket{\alpha,\boldsymbol\kappa},
\end{aligned}
\label{eq: residual_translation_field_matrix_element}
\end{equation}
where the field $\hat W_q^\mu(x)$ is the full Heisenberg interacting field.

\subsection{Covariance with respect to a reference}

The residual translations act within the hypersurfaces of constant plane-wave
phase,
\begin{equation}
    \Sigma_\phi
    =
    \left\{
        x\in\mathbb R^{1,3}
        \,\middle|\,
        n\cdot x=\phi
    \right\},
\end{equation}
such that, choosing inside each $\Sigma_\phi$ a reference point $x_\phi$, any
$x\in\Sigma_\phi$ can be decomposed as
\begin{equation}
    x=x_\phi+\xi_x,
    \qquad
    n\cdot x_\phi=\phi,
    \qquad
    \xi_x\in n^\perp,
\end{equation}
allowing Eq.~\eqref{eq: residual_translation_field_matrix_element} to be parametrized with respect to the chosen reference point, for both frequency sectors $r = \pm 1$,
\begin{equation}
\begin{aligned}
    \bra{\Omega}
    \hat W_q^\mu(x)
    \ket{\alpha,\boldsymbol\kappa}
    &=
    e^{-i\kappa(\xi_x)}
    F_{q,+,\alpha,\boldsymbol\kappa}^{\mu}(\phi),
    \\
    \bra{\alpha,\boldsymbol\kappa}
    \hat W_q^\mu(x)
    \ket{\Omega}
    &=
    e^{+i\kappa(\xi_x)}
    F_{q,-,\alpha,\boldsymbol\kappa}^{\mu}(\phi),
\end{aligned}
\label{eq: residual_translation_matrix_element}
\end{equation}
where, treating the two frequency sectors simultaneously,
\begin{equation}
\label{eq: covariant_element_fixed_phi}
\begin{aligned}
    F_{q,r,\alpha,\boldsymbol\kappa}^{\mu}(\phi)
    \coloneqq
    \begin{cases}
        \bra{\Omega}
        \hat W_q^\mu(x_\phi)
        \ket{\alpha, \boldsymbol{\kappa}},
        & r=+,
        \\[1mm]
        \bra{\alpha, \boldsymbol{\kappa}}
        \hat W_q^\mu(x_\phi)
        \ket{\Omega},
        & r=-.
    \end{cases}
\end{aligned}
\end{equation}
contains the remaining dependence on the phase $\phi$, not determined by the continuous residual translation symmetry.

\subsection{Wightman function and spectral kernel}

Resolving the identity in terms of these interaction states,
\begin{equation}
    \mathbf 1
    =
    \ket{\Omega}\bra{\Omega}
    +
    \sum_\alpha
    \int d\mu(\boldsymbol\kappa)\,
    \ket{\alpha,\boldsymbol\kappa}
    \bra{\alpha,\boldsymbol\kappa},
    \label{eq: residual_group_identity}
\end{equation}
where $d\mu(\boldsymbol\kappa)$ denotes the joint spectral measure
associated with the three commuting residual-translation generators, and inserting it into the positive-frequency function gives
\begin{equation}
\begin{aligned}
    \mathcal{G}_{q,+}^{\mu\nu}(x,x')
    &=
    \bra{\Omega}
    \hat W_q^\mu(x)
    \hat W_{-q}^\nu(x')
    \ket{\Omega}
    \\
    &=
    \sum_\alpha
    \int d\mu(\boldsymbol\kappa)\,
    \begin{aligned}[t]
        &\bra{\Omega}
        \hat W_q^\mu(x)
        \ket{\alpha,\boldsymbol\kappa}
        \times
        \\
        &\bra{\alpha,\boldsymbol\kappa}
        \hat W_{-q}^\nu(x')
        \ket{\Omega},
    \end{aligned}
\end{aligned}.
\label{eq: residual_KL_identity_insertion}
\end{equation}
where the contribution of the vacuum projector vanishes, since the
global electromagnetic $U(1)$ symmetry is unbroken and the vacuum is
a charge eigenstate, implying
\begin{equation}
\bra{\Omega}
\hat W_q^\mu(x)
\ket{\Omega}
=
0.
\end{equation}

Using Eq.~\eqref{eq: residual_translation_matrix_element} and its conjugate,
\begin{equation}
\begin{aligned}
    \mathcal{G}_{q,r}^{\mu\nu}(x,x')
    =
    \int d\mu(\boldsymbol\kappa)\,
    e^{-i r \kappa(\xi_x-\xi_{x'})}\,
    \rho_{q,r}^{\mu\nu}
    \left(
        \boldsymbol\kappa;\phi,\phi'
    \right),
\end{aligned}
\label{eq: residual_KL_spectral_representation}
\end{equation}
where the spectral kernel reads
\begin{equation}
    \rho_{q,r}^{\mu\nu}
    \left(
        \boldsymbol\kappa;\phi,\phi'
    \right)
    \coloneqq
    \sum_\alpha
    F_{q,r,\alpha,\boldsymbol\kappa}^{\mu}(\phi)\,
    \bar F_{q,r,\alpha,\boldsymbol\kappa}^{\nu}(\phi'),
\label{eq: residual_KL_spectral_kernel}
\end{equation}
associated with the continuous residual translation group. In contrast to the ordinary K\"all\'en--Lehmann spectral density, it depends not only on the three conserved momentum eigenvalues $\boldsymbol\kappa$, but also separately on the phases $\phi$ and $\phi'$. 

\subsection{Resolution with respect to the asymptotic spectral parameter \texorpdfstring{$s$}{s}}

The residual translation symmetry fixes only the three momentum
components contained in $\boldsymbol\kappa$. The remaining component
$p^-$ is not the eigenvalue of a conserved translation generator,
since translations along $\widetilde n^\mu$ change the plane-wave
phase. In the one-particle asymptotic Volkov--Proca theory, however,
the Ritus construction provides an additional spectral variable through
Eq.~\eqref{eq: Ritus_diagonalization_EOM_operator}. Introducing the
background-covariant squared-momentum operator
\begin{equation}
    (\mathcal K_q)^\alpha{}_\beta
    \coloneqq
    \Pi_q^2 \delta^\alpha{}_\beta
    -
    2iq\mathcal F^\alpha{}_\beta ,
\end{equation}
the Ritus modes satisfy
\begin{equation}
    (\mathcal K_q)^\alpha{}_\beta
    E_q^\beta{}_\gamma(p,x)
    =
    p^2 E_q^\alpha{}_\gamma(p,x).
\end{equation}
We therefore introduce $s \coloneqq p^2$ as the corresponding one-particle spectral parameter. The operator
$\mathcal K_q$ thus plays, for the asymptotic Volkov--Proca modes, the role played
by $P^2$ in the vacuum theory. At fixed $\boldsymbol\kappa$, the value of $s$ equivalently determines the
remaining spectral momentum component $p^-$.

For the full asymptotic Volkov Hilbert space, we assume that the one-particle spectral variable $s$ extends to the spectrum of a self-adjoint operator $\widehat S_{\mathrm V}$, whose restriction to the one-particle sector is represented by $\mathcal K_q$. We further assume that $\widehat S_{\mathrm V}$ strongly commutes with the residual-translation generators $\widehat Q_A^{\mathrm V}$, so that these operators admit a joint spectral resolution. The corresponding asymptotic states may therefore be chosen as simultaneous generalized eigenstates,
\begin{equation}
\begin{aligned}
\widehat S_{\mathrm V}
\ket{\beta,\boldsymbol\kappa,s}_{\mathrm V}
&=
s\,
\ket{\beta,\boldsymbol\kappa,s}_{\mathrm V},\\
\widehat Q_A^{\mathrm V}
\ket{\beta,\boldsymbol\kappa,s}_{\mathrm V}
&=
\kappa_A\,
\ket{\beta,\boldsymbol\kappa,s}_{\mathrm V},
\qquad
A=1,2,3.
\end{aligned}
\end{equation}
Here, $\ket{\beta,\boldsymbol\kappa,s}_{\mathrm V}$ denotes an asymptotic Volkov state labelled by the residual spectral variables $\boldsymbol\kappa$, the spectral parameter $s$, and the remaining discrete and continuous quantum numbers collected in $\beta$.

Assuming the existence\footnote{The existence and asymptotic completeness of the M\o ller operators are beyond the scope of this article and are therefore assumed.} of the M\o ller operators $\Omega_{\mathrm{in/out}}$ relating the Volkov asymptotic dynamics to the full interacting theory, the corresponding exact scattering states are defined by
\begin{equation}
\ket{\beta,\boldsymbol\kappa,s;\mathrm{in/out}}
\coloneqq
\Omega_{\mathrm{in/out}},
\ket{\beta,\boldsymbol\kappa,s}_{\mathrm V}.
\end{equation}
When the M\o ller operators preserve the residual translation symmetry, they intertwine the corresponding generators,
\begin{equation}
\widehat Q_A\,
\Omega_{\mathrm{in/out}}
=
\Omega_{\mathrm{in/out}}\,
\widehat Q_A^{\mathrm V},
\end{equation}
so that the residual spectral labels $\boldsymbol\kappa$ are preserved between the asymptotic and exact scattering states.

On the corresponding scattering subspace, we define
\begin{equation}
\widehat S_{\mathrm{in/out}}
\coloneqq
\Omega_{\mathrm{in/out}}\,
\widehat S_{\mathrm V}\,
\Omega_{\mathrm{in/out}}^\dagger .
\end{equation}
Using the isometric property of the M\o ller operators on the asymptotic Hilbert space, this gives
\begin{equation}
\widehat S_{\mathrm{in/out}}
\ket{\beta,\boldsymbol\kappa,s;\mathrm{in/out}}
=
s\,
\ket{\beta,\boldsymbol\kappa,s;\mathrm{in/out}} .
\end{equation}

The parameter $s$ is therefore inherited from the spectral
decomposition of the asymptotic Volkov theory and should not be
identified with the eigenvalue of the square of a conserved
four-momentum operator of the full interacting theory.

Assuming asymptotic completeness, the complete set of intermediate
states may consequently be resolved with respect to this spectral
parameter. The label $\alpha$ may therefore be decomposed as
\begin{equation}
    \alpha=(s,\beta),
\end{equation}
and the corresponding sum written schematically as
\begin{equation}
    \sum_\alpha
    \longrightarrow
    \int_0^\infty ds\,
    \sum_\beta .
\end{equation}
The $s$-resolved spectral kernel is then defined by
\begin{equation}
\begin{aligned}
    \varrho_{q,r}^{\mu\nu}
    \left(
        \boldsymbol\kappa,s;
        \phi,\phi'
    \right)
    \coloneqq{}&
    \sum_\beta
    F_{q,r,\beta,\boldsymbol\kappa,s}^{\mu}(\phi)\,
    \bar F_{q,r,\beta,\boldsymbol\kappa,s}^{\nu}(\phi').
\end{aligned}
\label{eq: s_resolved_spectral_kernel}
\end{equation}
The residual spectral kernel from
Eq.~\eqref{eq: residual_KL_spectral_kernel} is recovered by integrating
over the additional spectral parameter,
\begin{equation}
\begin{aligned}
    \rho_{q,r}^{\mu\nu}
    \left(
        \boldsymbol\kappa;
        \phi,\phi'
    \right)
    =
    \int_0^\infty ds\,
    \varrho_{q,r}^{\mu\nu}
    \left(
        \boldsymbol\kappa,s;
        \phi,\phi'
    \right).
\end{aligned}
\label{eq: s_integrated_spectral_kernel}
\end{equation}

In the one-particle Volkov--Proca sector, the non-negative spectral
support follows directly from the Ritus construction. The dressed
kinetic momentum satisfies
\begin{equation}
    \pi_q^\mu(p,\phi)
    =
    \Lambda_q^\mu{}_\nu(p,\phi)\,p^\nu ,
\end{equation}
where $\Lambda_q(p,\phi)$ is a proper orthochronous Lorentz
transformation. Hence
\begin{equation}
    \pi_q^2(p,\phi)
    =
    p^2
    =
    s,
    \qquad
    p\in\overline V_+
    \Longrightarrow
    \pi_q(p,\phi)\in\overline V_+ ,
\end{equation}
where the closed forward light cone is
\begin{equation}
    \overline V_+
    \coloneqq
    \left\{
        p\in\mathbb R^{1,3}
        \,\middle|\,
        p^2\geq 0,\;
        p^0\geq 0
    \right\} .
\end{equation}
The Volkov dressing therefore preserves both the causal character and the time orientation of the one-particle spectral momentum. In
particular, the physical one-particle contribution is supported at
$s=M_W^2$. For the full asymptotic spectral resolution, we impose the
corresponding spectral condition
\begin{equation}
    \sigma(\widehat S_{\mathrm V})
    \subseteq [0,\infty),
\end{equation}
which justifies taking the spectral integration over
$s\in[0,\infty)$.

\subsection{One-particle spectral amplitudes on the physical mass shell}

We now restrict the general spectral decomposition to the isolated
one-particle sector. In contrast to the generic spectral states
$\ket{\beta,\boldsymbol\kappa,s}_{\mathrm{V}}$, which need not admit an
interpretation in terms of asymptotic single-particle states, the
isolated physical $W$-boson contribution is supported on the mass shell $s = M_W^2$. This isolated sector is precisely the one associated with the asymptotic Volkov--Proca field defined by the LSZ condition in Eq.~\eqref{eq: LSZ_asymptotic_field_condition}, which is further assumed.

The pair that reads in light-cone coordinates
\begin{equation}
    \left(
        \boldsymbol\kappa,s
    \right)
    =
    \left(
        p^+,
        p^{\perp,1},
        p^{\perp,2},
        p^2
    \right)
\end{equation}
determines the complete momentum $p^\mu$, since the remaining
\begin{equation}
    p^-
    =
    \frac{
        s+\boldsymbol p_\perp^2
    }{
        2p^+
    },
    \qquad 
    p^+ \neq 0,
\end{equation}
is fixed by $p^2 = 2p^+p^- - \boldsymbol p_\perp^2$, where $\boldsymbol p_\perp^2 \coloneqq \sum_{i=1}^{2} \left( p^{\perp,i} \right)^2$. 

For the isolated one-particle contribution, however, we set
$s=M_W^2$, which defines $p^\mu \equiv p^\mu_{M_W^2}(\boldsymbol\kappa)$. The remaining one-particle quantum numbers are the charge sector $q=\pm1$ and the physical polarization label
$\lambda=1,2,3$. The corresponding asymptotic one-particle states are
defined by
\begin{subequations}
\label{eq: asymptotic_VP_one_particle_states}
\begin{align}
    \ket{
        q,\lambda,\boldsymbol p
    }_\mathrm{as}
    &\equiv
    \hat a_{q,\lambda}^{\mathrm{as}\dagger}
    (\boldsymbol p)
    \ket{\Omega},
    \\
    \ket{
        -q,\lambda,\boldsymbol p
    }_\mathrm{as}
    &\equiv
    \hat b_{-q,\lambda}^{\mathrm{as}\dagger}
    (\boldsymbol p)
    \ket{\Omega},
\end{align}
\end{subequations}
where $\mathrm{as}=\mathrm{in}$ or $\mathrm{out}$.

The vacuum--to--one-particle matrix elements of the
interacting Heisenberg field are related to those of the asymptotic
Volkov--Proca field by
\begin{subequations}
\label{eq: Heisenberg_asymptotic_one_particle_matrix_elements}
\begin{align}
    &\lim_{t\to t_{\mathrm{as}}}
    \bra{\Omega}
    \hat W_q^\mu(x)
    \ket{
        q,\lambda,\boldsymbol p
    }_{\mathrm{as}}
    \notag\\
    &\qquad =
    Z_W^{1/2}
    \bra{\Omega}
    \hat W_q^{\mu,\mathrm{as}}(x)
    \ket{
        q,\lambda,\boldsymbol p
    }_{\mathrm{as}},
    \\
    &\lim_{t\to t_{\mathrm{as}}}
    {}_{\mathrm{as}}\!\bra{
        -q,\lambda,\boldsymbol p
    }
    \hat W_q^\mu(x)
    \ket{\Omega}
    \notag\\
    &\qquad =
    Z_W^{1/2}
    {}_{\mathrm{as}}\!\bra{
        -q,\lambda,\boldsymbol p
    }
    \hat W_q^{\mu,\mathrm{as}}(x)
    \ket{\Omega},
\end{align}
\end{subequations}
with the time asymptotics given by
\begin{equation}
    t_{\mathrm{as}}
    \coloneqq
    \begin{cases}
        -\infty, & \mathrm{as}=\mathrm{in},\\
        +\infty, & \mathrm{as}=\mathrm{out},
    \end{cases}
\end{equation}
such that inserting in Eq.~\eqref{eq: covariant_element_fixed_phi} leads to
\begin{equation}
    F_{q,r,\lambda, \boldsymbol\kappa}^{\mathrm{1p}\,\mu}
    (\phi)
    =
    Z_W^{1/2}
    w_{q,\lambda}^{r\,\mu}
    (\boldsymbol p,x_\phi),
    \qquad
    p^2=M_W^2.
\label{eq: one_particle_spectral_amplitude_mode_relation}
\end{equation}

Using the residual-translation factorization,
\begin{equation}
    w_{q,\lambda}^{r\,\mu}
    (\boldsymbol p,x)
    =
    e^{-ir\kappa(\xi_x)}
    w_{q,\lambda}^{r\,\mu}
    (\boldsymbol p,x_\phi),
\end{equation}
the $1$-particle matrix element of the interacting field becomes
\begin{equation}
\begin{aligned}
    &
    \bra{\Omega}
    \hat W_{q}^\mu(x)
    \ket{
        q,\lambda,\boldsymbol p
    }_\mathrm{as}
    =
    Z_W^{1/2}
    e^{-i\kappa(\xi_x)}
    w_{q,\lambda}^{+\,\mu}
    (\boldsymbol p,x_\phi),
\end{aligned}
\end{equation}
with the analogous relation for the negative-frequency sector.

Choosing the reference point $x_\phi = \phi\,\widetilde n$, one has $p\cdot x_\phi = p^-\phi$.
Since the Volkov--Proca mode is
\begin{equation}
    w_{q,\lambda}^{r\,\mu}(p,x)
    =
    E_q^\mu{}_\rho(rp,x)\,
    \varepsilon_{\lambda}^\rho(rp),
\end{equation}
the one-particle spectral amplitude becomes
\begin{equation}
\begin{aligned}
    F_{q,r,\lambda,\boldsymbol\kappa}^{\mathrm{1p}\,\mu}
    (\phi)
    ={}&
    Z_W^{1/2}
    E_q^\mu{}_\rho
    (rp,x_\phi)\,
    \varepsilon_{(\lambda)}^\rho(rp)
    \\
    ={}&
    Z_W^{1/2}
    e^{-ir p^-\phi}
    E_q^\mu{}_\rho
    (rp,\phi)\,
    \varepsilon_{(\lambda)}^\rho(rp).
\end{aligned}
\label{eq: one_particle_spectral_amplitude_fixed_phase}
\end{equation}

The contribution of the isolated $1$-particle sector to the residual
spectral kernel follows by summing over the three physical
polarizations,
\begin{equation}
\begin{aligned}
    \left.
    \rho_{q,r}^{\mu\nu}
    \left(
        \boldsymbol\kappa;
        \phi,\phi'
    \right)
    \right|_{\mathrm{1p}}
    ={}&
    \sum_{\lambda=1}^{3}
    F_{q,r,\lambda}^{\mathrm{1p}\,\mu}
    (\boldsymbol\kappa;\phi)
    \,
    \bar F_{q,r,\lambda}^{\mathrm{1p}\,\nu}
    (\boldsymbol\kappa;\phi'),
\end{aligned}
\label{eq: one_particle_residual_spectral_kernel_definition}
\end{equation}
which, using Eq.~\eqref{eq: one_particle_spectral_amplitude_fixed_phase}, becomes
\begin{equation}
\begin{aligned}
    \left.
    \rho_{q,r}^{\mu\nu}
    \left(
        \boldsymbol\kappa;
        \phi,\phi'
    \right)
    \right|_{\mathrm{1p}}
    ={}&
    Z_W
    e^{-ir p^-(\phi-\phi')}
    \mathcal N^{\mu\nu}
    \left(
        rp;
        M_W^2;
        \phi,\phi'
    \right).
\end{aligned}
\label{eq: one_particle_residual_KL_Ritus_kernel}
\end{equation}

Inserting Eq.~\eqref{eq: one_particle_residual_KL_Ritus_kernel}
into Eq.~\eqref{eq: residual_KL_spectral_representation} and
combining the residual-translation and phase-dependent factors using
\begin{equation}
    p\cdot(x-x')
    =
    \kappa(\xi_x-\xi_{x'})
    +
    p^-(\phi-\phi'),
\end{equation}
the $1$-particle contribution to the Wightman function is
\begin{equation}
\begin{aligned}
    \left.
    \mathcal{G}_{q,r}^{\mu\nu}(x,x')
    \right|_{\mathrm{1p}}
    ={}&
    Z_W
    \int
    d\mu(\boldsymbol\kappa)\,
    e^{-ir p\cdot(x-x')}
    \\
    &\times
    \mathcal N^{\mu\nu}
    \left(
        rp;
        M_W^2;
        \phi,\phi'
    \right),
\end{aligned}
\label{eq: one_particle_Wightman_Ritus}
\end{equation}
readily plugged into the propagators of Eqs.~\eqref{eq: Feynman_propagator_Wightman} and \eqref{eq: retarded_advanced_Jordan}.\\

\subsection{Spectral support decomposition of the spectral function}

The on-shell condition in Eq.~\eqref{eq: on-shell_support} determines the support of the isolated $1$-particle contribution to the spectral function, so that the spectral function of the interacting field may be decomposed as
\begin{equation}
\begin{aligned}
    \varrho_{q,r}^{\mu\nu}
    \left(
        \boldsymbol\kappa,s;
        \phi,\phi'
    \right)
    ={}&
    Z_W\,
    \delta(s-M_W^2)\,
    \mathcal K_{q,r}^{(0)\,\mu\nu}
    \left(
        \boldsymbol\kappa;
        \phi,\phi'
    \right)
    \\
    &+
    \varrho_{q,r,\mathrm{cont}}^{\mu\nu}
    \left(
        \boldsymbol\kappa,s;
        \phi,\phi'
    \right),
\end{aligned}
\label{eq: spectral_kernel_support_decomposition}
\end{equation}
where $Z_W$ is the residue associated with the isolated physical
$1$-particle contribution, and we have introduced the corresponding
unit-residue Volkov--Proca kernel
\begin{equation}
\begin{aligned}
    \mathcal K_{q,r}^{(0)\,\mu\nu}
    \left(
        \boldsymbol\kappa;
        \phi,\phi'
    \right)
    \coloneqq{}&
    e^{-ir p^-(\phi-\phi')}\,
    \mathcal N^{\mu\nu}
    \left(
        rp;
        M_W^2;
        \phi,\phi'
    \right),
\end{aligned}
\label{eq: one_particle_unit_residue_kernel}
\end{equation} 
while $\varrho_{q,r,\mathrm{cont}}^{\mu\nu}$ contains the continuum contributions.

\section{Phase evolution of the Volkov--Proca modes}
\label{sec: phase_evolution}

We now consider the phase evolution of the isolated one-particle
contribution to the spectral amplitude. Throughout this section, the
spectral momentum is again restricted to the physical mass shell, with $p \equiv p_{M_W^2}(\boldsymbol\kappa)$.

Introducing the full-phase dependent Ritus matrix
\begin{equation}
    \mathcal E_{q,r}
    \left(
        p,\phi
    \right)
    \coloneqq
    e^{-ir p^-\phi}\,
    E_q
    \left(
        r p,\phi
    \right),
\label{eq: phase_Ritus_operator}
\end{equation}
the spectral amplitude from Eq.~\eqref{eq: one_particle_spectral_amplitude_fixed_phase} takes the form
\begin{equation}
\begin{aligned}
    F_{q,r,\lambda}^{\mathrm{1p}\,\mu}
    \left(
        p; \phi
    \right)
    ={}&
    Z_W^{1/2}\,
    \mathcal E_{q,r}
    \left(
        p,\phi
    \right)^\mu{}_\nu
    \varepsilon_{(\lambda)}^\nu(rp),
\end{aligned}
\label{eq: spectral_amplitude_phase_operator}
\end{equation}

The phase-evolution operator,
\begin{equation}
    \mathsf U_{q,r}^{\mathrm{ph}}
    \left(
        p;\phi,\phi_0
    \right)
    \coloneqq
    \mathcal E_{q,r}
    \left(
        p,\phi
    \right)
    \mathcal E_{q,r}^{-1}
    \left(
        p,\phi_0
    \right),
\label{eq: phase_evolution_operator}
\end{equation}
describes the evolution from hypersurface $\Sigma_{\phi_0}$ to
$\Sigma_\phi$, so that
\begin{equation}
    F_{q,r,\lambda}^{\mathrm{1p}\,\mu}
    \left(
        p; \phi
    \right)
    =
    \left[
        \mathsf U_{q,r}^{\mathrm{ph}}
        \left(
            p;\phi,\phi_0
        \right)
    \right]^\mu{}_\nu
    F_{q,r,\lambda}^{\mathrm{1p}\,\nu}
    \left(
        p; \phi_0
    \right).
\label{eq: phase_evolution_spectral_amplitude}
\end{equation}

By construction, the phase-evolution operator satisfies
\begin{equation}
    \mathsf U_{q,r}^{\mathrm{ph}}
    (p;\phi_0,\phi_0)
    =
    \mathbf 1,
\end{equation}
together with the composition and inverse relations
\begin{align}
    \mathsf U_{q,r}^{\mathrm{ph}}
    (p;\phi_2,\phi_1)\,
    \mathsf U_{q,r}^{\mathrm{ph}}
    (p;\phi_1,\phi_0)
    &=
    \mathsf U_{q,r}^{\mathrm{ph}}
    (p;\phi_2,\phi_0),
    \\
    \left[
        \mathsf U_{q,r}^{\mathrm{ph}}
        (p;\phi,\phi_0)
    \right]^{-1}
    &=
    \mathsf U_{q,r}^{\mathrm{ph}}
    (p;\phi_0,\phi).
\end{align}

The phase generator is defined by
\begin{equation}
    i\,\partial_\phi
    \mathsf U_{q,r}^{\mathrm{ph}}
    (p;\phi,\phi_0)
    =
    \mathsf H_{q,r}(p,\phi)\,
    \mathsf U_{q,r}^{\mathrm{ph}}
    (p;\phi,\phi_0),
\label{eq: phase_evolution_generator}
\end{equation}
which, using Eq.~\eqref{eq: phase_Ritus_operator}, can be extracted as
\begin{widetext}
\begin{equation}
\begin{aligned}
    \mathsf H_{q,r}(p,\phi)
    &=
    i\,
    \partial_\phi
    \mathcal E_{q,r}(p,\phi)\,
    \mathcal E_{q,r}^{-1}(p,\phi)
    \\
    &=
    r p^-\,\mathbf 1
    +
    i\,
    \partial_\phi
    E_q(r p,\phi)\,
    E_q^{-1}(r p,\phi)\\
    &=
    \left[
        r p^-
        +
        \partial_\phi S_q(r p,\phi)
    \right]\mathbf 1
    +
    i\,
    \partial_\phi
    \Lambda_q(r p,\phi)\,
    \Lambda_q^{-1}(r p,\phi),
\end{aligned}
\label{eq: phase_generator}
\end{equation}
\end{widetext}
where the last form makes explicit the split between the Volkov phase evolution and the spin-$1$ Lorentz transport.

Thus, at fixed residual momentum $\boldsymbol\kappa$, an on-shell
Volkov--Proca mode specified on a reference hypersurface
$\Sigma_{\phi_0}$, as in
Eq.~\eqref{eq: one_particle_spectral_amplitude_fixed_phase}, is
uniquely transported to any other constant-phase hypersurface
$\Sigma_\phi$ by
$\mathsf U_{q,r}^{\mathrm{ph}}(p;\phi,\phi_0)$, with
$p^2=M_W^2$.

\section{Summary and conclusions}
\label{conclusions}

In conclusion, we have derived orthogonality and completeness identities for the Ritus matrices and similar orthogonality and completeness relations for the Volkov--Proca modes. The orthogonality relations are written in terms of the sesquilinear indefinite inner product induced by the conserved current. 

We have also analyzed the continuous residual translation symmetries of the theory and quantized the associated conserved charges, showing that they admit diagonal representations in terms of the creation and annihilation operators. Consistently with the reduced translation symmetry of the plane-wave background, we have derived the corresponding K\"all\'en--Lehmann spectral representation for the Volkov--Proca field and constructed the phase-evolution operator governing the remaining dependence on the plane-wave phase.

Furthermore, by solving the Green equation, we have constructed the off-shell covariant and on-shell representations of the causal propagators, analyzed their analytic structure, and expressed the corresponding Wightman functions and projectors onto the positive- and negative-frequency subspaces. We have also derived a representation of the Feynman Volkov--Proca propagator in terms of the scalar Volkov propagator, in which the background-induced spin-one dressing is encoded in the bilocal Lorentz transport, while the longitudinal sector retains the corresponding dressed Proca structure.

\section*{Acknowledgments}

The author is grateful to Emiliano Manuel Cancino, Dr. Nyx Shiva, Dr. Victor Ambruș, Dr. Cosmin Crucean and Dr. Maxim Chernodub for useful comments on the manuscript. 

This work was funded by the EU’s NextGenerationEU instrument through the National Recovery and Resilience Plan of Romania - Pillar III-C9-I8, managed by the Ministry of Research, Innovation and Digitization, within the project entitled ``Facets of Rotating Quark-Gluon Plasma'' (FORQ), contract no.~760079/23.05.2023 code CF 103/15.11.2022. 

\section*{Appendices}

Throughout the proofs, we employ the light-cone basis
\begin{subequations}
\label{eq: light_cone_basis}
\begin{equation}
    n^\mu=(1,\boldsymbol n),
    \qquad
    \tilde n^\mu=\frac{1}{2}(1,-\boldsymbol n),
\label{eq: light_cone_basis_null}
\end{equation}
with $n^2=\tilde n^2=0$ and $n\cdot\tilde n=1$, completed by
\begin{equation}
    a_j^\mu=(0,\boldsymbol a_j),
    \qquad
    j=1,2,
\label{eq: light_cone_basis_transverse}
\end{equation}
\end{subequations}
satisfying the transversality and normalization conditions
\begin{equation}
    n\cdot a_j
    =
    \tilde n\cdot a_j
    =
    0,
    \qquad
    a_j\cdot a_{j'}
    =
    -\delta_{jj'}.
\end{equation}

For an arbitrary vector $x^\mu$, we define the corresponding light-cone coordinates by projecting onto these basis vectors,
\begin{equation}
\begin{aligned}
    x^+ &\equiv n\cdot x,
    \qquad
    x^- \equiv \tilde n\cdot x,
    \\
    x^{\perp,j} &\equiv -a_j\cdot x,
    \qquad
    j=1,2.
\end{aligned}
\label{eq: light_cone_coordinates}
\end{equation}

\appendix

\section{Gauge freedom of the background}
\label{app: gauge_choice}

In the light-cone basis of Eq.~\eqref{eq: light_cone_basis}, the vector potential is
\begin{equation}
    \mathcal{A}^\mu
    =
    (n\cdot\mathcal{A})\tilde n^\mu
    +
    (\tilde n\cdot\mathcal{A})n^\mu
    +
    \mathcal{A}_\perp^\mu.
\end{equation}
with $\mathcal{A}^+ \equiv n\cdot\mathcal{A}$ and $\mathcal{A}^- \equiv \tilde n\cdot\mathcal{A}$, as in Eq.~\eqref{eq: light_cone_coordinates}.

The Lorenz gauge condition for $\mathcal{A}^\mu=\mathcal{A}^\mu(\phi)$ reads
\begin{equation}
    \partial_\mu \mathcal{A}^\mu
    =
    n_\mu \mathcal{A}^{\mu\prime}
    =
    (n\cdot\mathcal{A})^\prime
    =
    0 \implies n\cdot\mathcal{A}=C,
\end{equation}
constant which can be removed by a gauge transformation
\begin{equation}
    \mathcal{A}^\mu
    \longrightarrow
    \mathcal{A}^\mu+\partial^\mu\Lambda,
\end{equation}
with $\Lambda=\Lambda(\tilde\phi)$ with $\tilde\phi=\tilde n \cdot x$, from which follows
\begin{equation}
    \partial^\mu\Lambda
    =
    \tilde n^\mu \Lambda^\prime(\tilde\phi),
\end{equation}
so that
\begin{equation}
    n\cdot\mathcal{A}
    \longrightarrow
    n\cdot\mathcal{A}
    +
    (n\cdot\tilde n)\Lambda^\prime
    =
    n\cdot\mathcal{A}
    +
    \Lambda^\prime.
\end{equation}

Choosing $\Lambda^\prime=-C$, or equivalently
\begin{equation}
    \Lambda=-C\tilde\phi,
\end{equation}
eliminates the $n-$light-cone component,
\begin{equation}
    \mathcal{A}^+
    =
    n\cdot\mathcal{A}
    =
    0.
\end{equation}

This transformation preserves the Lorenz gauge since
\begin{equation}
    \Box\Lambda(\tilde\phi)
    =
    \tilde n^2\Lambda^{\prime\prime}(\tilde\phi)
    =
    0.
\end{equation}

After imposing $\mathcal{A}^+=0$, the potential becomes
\begin{equation}
    \mathcal{A}^\mu
    =
    \mathcal{A}^- n^\mu
    +
    \mathcal{A}_\perp^\mu.
\end{equation}

There remains the residual Lorenz gauge freedom with an arbitrary function
$\Lambda=\Lambda(\phi)$. In this case,
\begin{equation}
    \partial^\mu\Lambda(\phi)
    =
    n^\mu\Lambda^\prime(\phi),
\end{equation}
and therefore
\begin{equation}
    \mathcal{A}^\mu(\phi)
    \longrightarrow
    \mathcal{A}^\mu(\phi)
    +
    n^\mu\Lambda^\prime(\phi),
\end{equation}
which again preserves the Lorenz gauge because
\begin{equation}
    \Box\Lambda(\phi)
    =
    n^2\Lambda^{\prime\prime}(\phi)
    =
    0.
\end{equation}

Moreover,
\begin{equation}
    n\cdot\mathcal{A}
    \longrightarrow
    n\cdot\mathcal{A}
    +
    n^2\Lambda^\prime
    =
    n\cdot\mathcal{A},
\end{equation}
so the condition $\mathcal{A}^+=0$ remains unchanged, whereas
\begin{equation}
    \mathcal{A}^-
    =
    \tilde n\cdot\mathcal{A}
    \longrightarrow
    \mathcal{A}^-
    +
    (\tilde n\cdot n)\Lambda^\prime
    =
    \mathcal{A}^-
    +
    \Lambda^\prime.
\end{equation}

Hence, choosing $\Lambda^\prime(\phi) = - \mathcal{A}^-(\phi)$, or equivalently,
\begin{equation}
    \Lambda(\phi)
    =
    -\int^\phi d\varphi\,\mathcal{A}^-(\varphi),
\end{equation}
eliminates the $\tilde{n}-$light-cone component,
\begin{equation}
    \mathcal{A}^-
    =
    \tilde n\cdot\mathcal{A}
    =
    0.
\end{equation}

The plane-wave can be chosen purely transverse,
\begin{equation}
    \mathcal{A}^\mu(\phi)
    =
    \mathcal{A}_\perp^\mu(\phi),
\end{equation}
with $n\cdot\mathcal{A} = \tilde n\cdot\mathcal{A} = 0$, or equivalently, 
\begin{equation}
    \label{eq: gauge_choice}
    \mathcal{A}^0 = 0, \qquad \boldsymbol{n} \cdot \boldsymbol{\mathcal{A}} = 0,
\end{equation}
for the wave vector $n^\mu = (1, \boldsymbol{n})$.\\

\section{Proofs of Ritus matrices properties}
\label{app: Ritus_properties}
The identities in Eq.~\eqref{eq: spectral_relations_Ritus}, together
with the diagonalization of the EOM operator in
Eq.~\eqref{eq: Ritus_diagonalization_EOM_operator}, constitute central
results used throughout the present work, in particular in
Secs.~\ref{sec: Volkov-Proca_propagator} and
\ref{sec: completenes_orthonormality_states}. We first compute the
action of the dressed momentum operator defined in
Eq.~\eqref{def: momentum_operator} on the Ritus matrix,
\begin{widetext}
\begin{equation}
\begin{aligned}
    \Pi_q^\alpha(\phi) E_{\mu\nu}(p,x)
    &=
    \left[
        p^\alpha
        +
        n^\alpha
        \left(
            \frac{q\,p\cdot\mathcal A(\phi)}{n\cdot p}
            -
            \frac{\mathcal A^2(\phi)}
                 {2\,n\cdot p}
        \right)
        -
        q\mathcal A^\alpha(\phi)
    \right]
    E_{\mu\nu}(p,x)
    +
    i n^\alpha
    \frac{q}{n\cdot p}
    \mathcal F_{\mu\rho}(\phi)
    E^\rho{}_\nu(p,x)
    \\
    &=
    \pi_q^\alpha(p,\phi)E_{\mu\nu}(p,x)
    +
    i n^\alpha
    \frac{q}{n\cdot p}
    \mathcal F_{\mu\rho}(\phi)
    E^\rho{}_\nu(p,x),
\end{aligned}
\label{eq: EOMDiag1}
\end{equation}
\end{widetext}
where $\pi_q^\mu(p,\phi)$ is the background-dressed momentum defined in
Eq.~\eqref{eq: dressed_momentum_contracted}. Acting once more with the dressed momentum,

\begin{widetext}
\begin{equation}
\begin{aligned}
    \Pi_q^2 (\phi)\, E_{\mu\nu}
    &=
    \Pi_{q,\alpha}
    \left[
        \pi_q^\alpha(p,\phi) E_{\mu\nu}
        +
        i n^\alpha
        \frac{q}{n\cdot p}
        \mathcal F_{\mu\rho}
        E^\rho{}_\nu
    \right]
    \\
    &=
    i\left(
        \partial_\alpha
        \pi_q^\alpha(p,\phi)
    \right)E_{\mu\nu}
    +
    \pi_q^\alpha(p,\phi)
    \Pi_{q,\alpha}E_{\mu\nu}
    +
    i\frac{q}{n\cdot p}
    n^\alpha
    \Pi_{q,\alpha}
    \left(
        \mathcal F_{\mu\rho}
        E^\rho{}_\nu
    \right).
\end{aligned}
\end{equation}
\end{widetext}

Since the derivative acts on the dressed momentum as
\begin{equation}
    \partial_\alpha
    \pi_q^\alpha(p,\phi)
    =
    n_\alpha
    \frac{d\pi_q^\alpha(p,\phi)}{d\phi}
    =
    \frac{d}{d\phi}
    \left(
        n\cdot\pi_q(p,\phi)
    \right)
    =
    0,
\end{equation}
and the wave vector undresses the momentum,
\begin{equation}
    n\cdot\pi_q(p,\phi)
    =
    n\cdot p,
    \qquad
    \pi_q^2(p,\phi)
    =
    p^2,
\end{equation}
the first nonvanishing term becomes
\begin{equation}
\begin{aligned}
    \pi_q^\alpha(p,\phi)
    \Pi_{q,\alpha}E_{\mu\nu}
    &=
    \pi_q^2(p,\phi)E_{\mu\nu}
    +
    i\frac{
        q\,\pi_q(p,\phi)\cdot n
    }{n\cdot p}
    \mathcal F_{\mu\rho}
    E^\rho{}_\nu
    \\
    &=
    p^2E_{\mu\nu}
    +
    iq\mathcal F_{\mu\rho}
    E^\rho{}_\nu.
\end{aligned}
\end{equation}
Moreover, using $n^2=0$, $n\cdot\mathcal A=0$, one has
\begin{equation}
    n^\alpha\partial_\alpha
    \mathcal F_{\mu\nu}(\phi)
    =
    n^2\mathcal F'_{\mu\nu}(\phi)
    =
    0,
\end{equation}
together with
\begin{equation}
    n^\alpha
    \Pi_{q,\alpha}
    E^\rho{}_\nu(p,x)
    =
    (n\cdot p)
    E^\rho{}_\nu(p,x).
\end{equation}
Applying the dressed momentum to the product gives
\begin{equation}
\begin{aligned}
    n^\alpha
    \Pi_{q,\alpha}
    \left(
        \mathcal F_{\mu\rho}
        E^\rho{}_\nu
    \right)
    &=
    i n^\alpha
    \left(
        \partial_\alpha\mathcal F_{\mu\rho}
    \right)
    E^\rho{}_\nu
    +
    \mathcal F_{\mu\rho}
    n^\alpha
    \Pi_{q,\alpha}
    E^\rho{}_\nu
    \\
    &=
    (n\cdot p)\,
    \mathcal F_{\mu\rho}
    E^\rho{}_\nu .
\end{aligned}
\end{equation}
Consequently,
\begin{equation}
\begin{aligned}
    i\frac{q}{n\cdot p}
    n^\alpha
    \Pi_{q,\alpha}
    \left(
        \mathcal F_{\mu\rho}
        E^\rho{}_\nu
    \right)
    &=
    i\frac{q}{n\cdot p}
    (n\cdot p)\,
    \mathcal F_{\mu\rho}
    E^\rho{}_\nu
    \\
    &=
    iq\mathcal F_{\mu\rho}
    E^\rho{}_\nu .
\end{aligned}
\end{equation}
Therefore, the EOM operator is diagonalized according to
\begin{equation}
    \left[
        \Pi_q^2(\phi)\eta_{\mu\rho}
        -
        2iq\mathcal F_{\mu\rho}(\phi)
    \right]
    E^\rho{}_\nu(p,x)
    =
    p^2E_{\mu\nu}(p,x).
    \label{eq: Ritus_diagonalization_explicit}
\end{equation}

Contracting in Eq.~\eqref{eq: EOMDiag1} and using
$n^\mu\mathcal F_{\mu\rho}=0$ yields
\begin{equation}
    \Pi_q^\mu(\phi)E_{\mu\nu}(p,x)
    =
    \pi_q^\mu(p,\phi)
    E_{\mu\nu}(p,x),
\end{equation}

The second relation in Eq.~\eqref{eq: dressed_momentum_general} follows analogously.

\section{Dressing and undressing of the momentum}
\label{app: momentum_dressing}

An important property of the Lorentz transformation
$\Lambda^\mu{}_{\nu}(p, \phi)$ entering the Volkov--Proca Ritus matrix is
that it maps the free momentum $p^\mu$ into the background-dressed
kinetic momentum $\pi^\mu(p, \phi)$, and conversely maps the latter back
to $p^\mu$. Explicitly,
\begin{subequations}
\label{eq: momentum_dressing_undressing}
\begin{align}
    \pi^\mu(p,\phi)\,
    \Lambda_{\mu}{}^{\alpha}(p,\phi)
    &=
    p^\alpha,
    \label{eq: momentum_undressing}
    \\
    \Lambda^{\mu}{}_{\alpha}(p,\phi)\,
    p^\alpha
    &=
    \pi^\mu (p,\phi).
    \label{eq: momentum_dressing}
\end{align}
\end{subequations}
Hence, $\Lambda(p,\phi)$ provides the Lorentz transformation relating the
constant asymptotic momentum $p^\mu$ to the local kinetic momentum
$\pi^\mu(p, \phi)$ along the Volkov trajectory.

\section{Phase transport of the dressed momentum}
\label{app: phase_transport_dressed_momentum}

It is useful to introduce the
relative Lorentz transformation
\begin{equation}
    \Lambda_q(p;\phi,\phi')
    \coloneqq
    \Lambda_q(p,\phi)\,
    \Lambda_q^{-1}(p,\phi'),
    \label{app: relative_Lambda_definition}
\end{equation}
between two phases $\phi'$ and $\phi$.

This transformation provides the Lorentz transport of the dressed momentum between two wavefronts of the background. Using Eq.~\eqref{eq: momentum_dressing}, then
\begin{equation}
\begin{aligned}
    &\Lambda_q(p;\phi,\phi')^\mu{}_\nu\,
    \pi_q^\nu(p,\phi')\\
    &\quad=
    \Lambda_q(p,\phi)^\mu{}_\alpha
    \left[
        \Lambda_q^{-1}(p,\phi')
    \right]^\alpha{}_\nu
    \Lambda_q(p,\phi')^\nu{}_\beta
    p^\beta
    \\
    &\quad=
    \Lambda_q(p,\phi)^\mu{}_\beta p^\beta
    =
    \pi_q^\mu(p,\phi).
\end{aligned}
\label{eq: appendix_dressed_momentum_transport}
\end{equation}

The Lorentz dressing entering the Ritus matrix is
\begin{equation}
    \Lambda_q(p; \phi)
    =
    \exp\left[
        \frac{q}{n\cdot p}\,
        N(\phi)
    \right],
    \qquad
    n\cdot p\neq0,
    \label{eq: appendix_local_Lorentz_dressing}
\end{equation}
with the phase-dependent null-rotation generator defined in Eq.~\eqref{eq: phase_dependent_null_rotation_generator}. Similarly, introducing
\begin{equation}
\begin{aligned}
    \Delta\mathcal A^\mu(\phi,\phi')
    &\coloneqq
    \mathcal A^\mu(\phi)
    -
    \mathcal A^\mu(\phi'),
    \\
    N(\Delta\mathcal A)^\mu{}_\nu
    &\coloneqq
    n^\mu\Delta\mathcal A_\nu
    -
    \Delta\mathcal A^\mu n_\nu,
    \end{aligned}
\end{equation}
the relative Lorentz transformation becomes
\begin{equation}
    \Lambda_q(p;\phi,\phi')
    =
    \exp\left[
        \frac{q}{n\cdot p}
        N\!\left(
            \Delta\mathcal A
        \right)
    \right],
    \label{eq: appendix_relative_Lambda_exponential}
\end{equation}
where it has been used that $\left[N(\phi), N(\phi')\right] = 0$. Since $N(\Delta\mathcal A)^3=0$, the exponential truncates exactly,
\begin{equation}
\begin{aligned}
    \Lambda_q(p;\phi,\phi')^\mu{}_\nu
    =
    \delta^\mu{}_\nu
    &+
    \frac{q}{n\cdot p}
    \left(
        n^\mu\Delta\mathcal A_\nu
        -
        \Delta\mathcal A^\mu n_\nu
    \right)
    \\
    &-
    \frac{q^2\Delta\mathcal A^2}
    {2(n\cdot p)^2}
    n^\mu n_\nu .
\end{aligned}
\label{eq: appendix_relative_Lambda_explicit}
\end{equation}

They are Lorentz transformations, since
\begin{equation}
    \Lambda_q(p;\phi,\phi')^\mu{}_\alpha\,
    \eta^{\alpha\beta}\,
    \Lambda_q(p;\phi,\phi')^\nu{}_\beta
    =
    \eta^{\mu\nu}.
    \label{eq: appendix_relative_Lambda_Lorentz}
\end{equation}
and also elements of the little group,
\begin{equation}
    \Lambda_q(p;\phi,\phi')^\mu{}_\nu n^\nu
    =
    n^\mu,
    \label{eq: appendix_relative_Lambda_n_invariant}
\end{equation}
keeping fixed the null wave vector $n^\mu$.

The relative transformations satisfy the relations,
\begin{subequations}
\label{eq: appendix_relative_Lambda_group_properties}
\begin{align}
    \Lambda_q(p;\phi,\phi)
    &=
    \mathbb 1,
    \\
    \Lambda_q^{-1}(p;\phi,\phi')
    &=
    \Lambda_q(p;\phi',\phi),
    \\
    \Lambda_q(p;\phi_1,\phi_2)\,
    \Lambda_q(p;\phi_2,\phi_3)
    &=
    \Lambda_q(p;\phi_1,\phi_3).
\end{align}
\end{subequations}

\section{Completeness and orthonormality of the Ritus matrices} 
\label{app: completeness_ortho_Ritus}

A detailed derivation of the completeness and orthonormality relations of Eq. (\ref{eq: Ritus_completeness}) and Eq. (\ref{eq: Ritus_orthogonality}), respectively, is provided, using integration over the light-cone coordinates, like in Ref.~\cite{Ritus1972, Seipt:2012nad}, where similar relations are obtained for Volkov fermions.

\subsection{Ritus completeness relation}

For the completeness relation, notice that the integral measure in light-cone coordinates is
\begin{equation*}
d^{4}p = dp^{+} dp^{-} d^{2}\boldsymbol{p}^{\perp},
\end{equation*}
and that the phase-dependent Ritus matrices\footnote{Dropping the subscript $q$ since the section refers to the same-charge sector.}
\begin{equation*}
    E^\mu_{\, \alpha} (p, \phi) = U (p, \phi)\, \Lambda^{\mu}_{\; \alpha} (p, \phi)
\end{equation*}
are independent of $p^{-}$, since $\mathcal{A}^{+}(\phi) = 0$. Consequently, the $p^-$ integration can be performed as
\begin{equation}
\label{eq: RItus_completeness_proof}
\begin{aligned}
& \int \frac{d^4 p}{(2\pi)^4}\, E^\mu_{\, \alpha} (p, x)\, \eta^{\alpha\beta}\, \bar{E}^\nu_{\, \beta} (p, x') = \delta \left(x^{+} - x^{\prime +}\right)\\
&\quad \times \int \frac{dp^{+} d^{2}\boldsymbol{p}^{\perp}}{(2 \pi)^{3}}\, e^{-i p^{+} \left( x^{-} - x^{\prime -} \right) + i \boldsymbol{p}^{\perp} \cdot (\boldsymbol{x}^{\perp} - \boldsymbol{x}^{\prime\perp})}\\
&\quad \times E^\mu_{\, \alpha} \left(p^{+}, \boldsymbol{p}^{\perp}, x^{+}\right) \eta^{\alpha \beta} \bar{E}^\nu_{\, \beta} \left( p^{+}, \boldsymbol{p}^{\perp}, x^{+}\right).
\end{aligned}
\end{equation}

The support of the Dirac delta distribution restricts the remaining
kernel to the equal-phase hypersurface $x^{+}=x^{\prime +}$. Using the
same-phase Lorentz relation
\begin{equation}
\label{eq: Lorentz_relation_same_phase}
    \Lambda^{\mu}_{\;\alpha}(p,\phi)\,
    \eta^{\alpha\beta}\,
    \Lambda^{\nu}_{\;\beta}(p,\phi)
    =
    \eta^{\mu\nu},
\end{equation}
together with the relation for the phase factors
\begin{equation}
    U(p, \phi)\,U^{*}(p, \phi)=1,
\end{equation}
the Ritus bilinear reduces at equal phase to
\begin{equation}
    \label{eq: same_point_Ritus_bilinear}
    E^\mu_{\;\alpha}
    (p^+,\boldsymbol p^\perp,x^+)\,
    \eta^{\alpha\beta}\,
    \bar E^\nu_{\;\beta}
    (p^+,\boldsymbol p^\perp,x^+)
    =
    \eta^{\mu\nu}.
\end{equation}

Consequently, Eq.~\eqref{eq: RItus_completeness_proof} becomes
\begin{equation}
\begin{aligned}
    &\int\frac{d^4p}{(2\pi)^4}\,
    E^\mu_{\;\alpha}(p,x)\,
    \eta^{\alpha\beta}\,
    \bar E^\nu_{\;\beta}(p,x') =
    \eta^{\mu\nu}
    \delta\left(x^+-x^{\prime+}\right)\\
    &\quad\times
    \int
    \frac{dp^+\,d^2\boldsymbol p^\perp}{(2\pi)^3}\,
    e^{-ip^+(x^--x^{\prime-})}
    e^{i\boldsymbol p^\perp\cdot
    (\boldsymbol x^\perp-\boldsymbol x^{\prime\perp})}.
\end{aligned}
\end{equation}
which, after integrating over $p^{+}$\footnote{The Ritus matrices are singular at the light-front zero mode $p^{+} = 0$.\\Nevertheless, the completeness relation can be understood as a distributional limit in which a neighborhood of $p^{+}=0$ is temporarily excluded.} and $\boldsymbol{p}^{\perp}$, leads to
\begin{equation}
\int \frac{d^4 p}{(2\pi)^4}\, E^\mu_{\, \alpha} (p, x)\, \eta^{\alpha\beta}\, \bar{E}^\nu_{\, \beta} (p, x') = \eta^{\mu \nu} \delta^{(4)} \left( x - x^{\prime} \right).
\end{equation}

\subsection{Ritus orthonormality relation}

To prove the orthonormality relation an analogous approach is followed. The integral measure in light-cone coordinates is 
\begin{equation*}
d^{4}x = dx^{+} dx^{-} d^{2}\boldsymbol{x}^{\perp},
\end{equation*}
such that, using that the phase-dependent Ritus matrices are independent of $x^{-}$, allows performing the $x^--$integral,
\begin{equation}
\begin{aligned}
\label{eq: Ritus_orthonormality_proof}
&\int d^{4}x \; E^{\mu}_{\; \alpha}(p,x)\, \eta^{\alpha \beta}\, \bar{E}^{\nu}_{\; \beta}(p^{\prime}, x) =  (2 \pi)\, \delta\left( p^{+} - p^{\prime +} \right)\\
&\quad \times \int dx^{+} d^{2}\boldsymbol{x}^{\perp}\, e^{-i x^{+} \left( p^{-} - p^{\prime -} \right) + i \boldsymbol{x}^{\perp} \cdot \left( \boldsymbol{p}^{\perp} - \boldsymbol{p}^{\prime\perp} \right)}\\
&\quad \times E^{\mu}_{\; \alpha}\left(p^+, \boldsymbol{p}^{\perp}, x^{+}\right) \eta^{\alpha \beta} \bar{E}^{\nu}_{\; \beta} \left(p^+, \boldsymbol{p}^{\prime\perp}, x^{+}\right).
\end{aligned}
\end{equation}
\vspace{0.2\baselineskip}

Since the phase-dependent Ritus matrices are independent of $\boldsymbol{x}^{\perp}$, the integral can be performed, yielding a Dirac delta distribution that restricts the Ritus matrices to $\boldsymbol{p}^{\perp} = \boldsymbol{p}^{\prime\perp}$, such that the phase-dependent Ritus matrices are evaluated at the same points. As before, using the same point Ritus bilinear from Eq.~\eqref{eq: same_point_Ritus_bilinear}, Eq. (\ref{eq: Ritus_orthonormality_proof}) simplifies to the result
\begin{equation}
\int d^{4}x \; E^{\mu}_{\; \alpha}(p,x)\, \eta^{\alpha \beta}\, \bar{E}^{\nu}_{\; \beta}(p^{\prime}, x) = \eta^{\mu \nu} (2 \pi)^{4} \delta^{(4)} \left( p - p^{\prime} \right).
\end{equation}

\section{Longitudinal contribution to the Green function}

The full vector Ritus matrix can be written as
\begin{equation}
    E_q(p,x)^\mu{}_\nu
    =
    \mathscr E_q(p,x)\,
    \Lambda_q(p,\phi)^\mu{}_\nu,
\end{equation}
in terms of the scalar Ritus factor from Eq.~\eqref{eq: scalar_Ritus_factor}. 
\vspace{-0.05\baselineskip}

The action of the dressed momentum operator on the scalar Ritus factor takes the particularly simple form
\vspace{0.5\baselineskip}
\begin{widetext}
\begin{equation}
\begin{aligned}
    \Pi_q^\alpha(\phi)\, \mathscr E_q(p,x)
    &=
    \left(
        i\partial^\alpha
        -
        q\mathcal A^\alpha(\phi)
    \right)
    \left[
        e^{-ip\cdot x}U_q(p,\phi)
    \right]
    \\
    &=
    e^{-ip\cdot x}
    \left[
        p^\alpha
        -
        q\mathcal A^\alpha(\phi)
        +
        n^\alpha
        \left(
            \frac{q\,p\cdot\mathcal A(\phi)}{n\cdot p}
            -
            \frac{\mathcal A^2(\phi)}
                 {2\,n\cdot p}
        \right)
    \right]
    U_q(p,\phi)
    \\
    &=
    \pi_q^\alpha(p,\phi)\,
    \mathscr E_q(p,x).
\end{aligned}
\label{eq: scalar_Ritus_spectral_relation}
\end{equation}
\end{widetext}
Thus, the scalar Ritus factor is an eigenfunction of the dressed
momentum operator, with the background-dressed momentum
$\pi_q^\alpha(p,\phi)$ as its local eigenvalue.

Inserting the completeness relation of Eq.~\eqref{eq: Ritus_completeness} into the right hand side of the transversality constraint of the Green function from Eq.~\eqref{eq: transversality_constraint_Green_function} gives
\begin{widetext}
\begin{equation}
\begin{aligned}
    \Pi_{q,\beta}(\phi)\,
    G_q^{\beta\gamma}(x,x')
    &=
    \frac{1}{M_W^2}
    \Pi_{q,\beta}(\phi)\,
    \eta^{\beta\gamma}\,
    \delta^{(4)}(x-x')
    \\
    &=
    \frac{1}{M_W^2}
    \int\frac{d^4p}{(2\pi)^4}\,
    \Pi_{q,\beta}(\phi)\,
    E_q^\beta{}_\rho(p,x)\,
    \eta^{\rho\sigma}\,
    \bar E_q^\gamma{}_\sigma(p,x')
    \\
    &=
    \frac{1}{M_W^2}
    \int\frac{d^4p}{(2\pi)^4}\,
    \pi_{q,\beta}(p,\phi)\,
    E_q^\beta{}_\rho(p,x)\,
    \eta^{\rho\sigma}\,
    \bar E_q^\gamma{}_\sigma(p,x').
\end{aligned}
\label{eq: Green_longitudinal_first_derivative}
\end{equation}
\end{widetext}

Using the undressing relation in Eq.~\eqref{eq: momentum_undressing}, the dressed momentum converts the Ritus matrix into the scalar Ritus factor multiplying the undressed momentum, as
\begin{equation}
\begin{aligned}
    \pi_{q,\beta}(p,\phi)
    E_q^\beta{}_\rho(p,x)
    &=
    \mathscr E_q(p,x)\,
    \pi_{q,\beta}(p,\phi)
    \Lambda_q^\beta{}_\rho(p,\phi)
    \\
    &=
    \mathscr E_q(p,x)\,p_\rho,
\end{aligned}
\end{equation}
such that the previous expression reduces to
\begin{equation}
    \Pi_{q,\beta}(\phi)
    G_q^{\beta\gamma}(x,x')
    =
    \frac{1}{M_W^2}
    \int\frac{d^4p}{(2\pi)^4}\,
    \mathscr E_q(p,x)\,
    p^\sigma\,
    \bar E_q^\gamma{}_\sigma(p,x').
    \label{eq: Green_longitudinal_scalar_Ritus}
\end{equation}
Acting once more with the dressed momentum operator and using
Eq.~\eqref{eq: scalar_Ritus_spectral_relation} and then using the dressing relation in Eq.~\eqref{eq: momentum_dressing} for
\begin{equation}
\begin{aligned}
    \pi_q^\alpha(p,\phi)
    \mathscr E_q(p,x)
    &=
    \mathscr E_q(p,x)
    \Lambda_q^\alpha{}_\rho(p,\phi)p^\rho
    \\
    &=
    E_q^\alpha{}_\rho(p,x)p^\rho,
\end{aligned}
\end{equation}
leads to the longitudinal contribution
\begin{widetext}
\begin{equation}
    \Pi_q^\alpha(\phi_x)\,
    \Pi_{q,\beta}(\phi_x)\,
    G_q^{\beta\gamma}(x,x')
    =
    \frac{1}{M_W^2}
    \int\frac{d^4p}{(2\pi)^4}\,
    E_q^\alpha{}_\rho(p,x)\,
    p^\rho p^\sigma\,
    \bar E_q^\gamma{}_\sigma(p,x').
\label{eq: Green_longitudinal_second_derivative}
\end{equation}
\end{widetext}

Combining this result with the completeness relation, the transverse
part of the EOM operator acting on the Green function becomes
\begin{equation}
\begin{aligned}
    &
    \left[
        \bigl(\Pi_q^2(\phi_x) - M^2_W\bigr)
        \delta^\alpha{}_\beta
        -
        2iq\mathcal F^\alpha{}_\beta(\phi_x)
    \right]
    G_q^{\beta\gamma}(x,x')
    \\
    &\qquad=
    \int\frac{d^4p}{(2\pi)^4}\,
    E_q^\alpha{}_\rho(p,x)
    \left(
        -\eta^{\rho\sigma}
        +
        \frac{p^\rho p^\sigma}{M_W^2}
    \right)
    \bar E_q^\gamma{}_\sigma(p,x').
\end{aligned}
\label{eq: Green_transverse_operator}
\end{equation}
Here, the longitudinal contribution has been transferred to the
right-hand side, while the term proportional to
$-\eta^{\rho\sigma}$ follows directly from the Ritus completeness
relation from Eq.~\eqref{eq: Ritus_completeness}.

\bibliography{bibliography}{}
\bibliographystyle{apsrev4-1}

\end{document}